\documentclass[iop]{emulateapj}
\usepackage{epstopdf}
\usepackage{graphicx}

\shorttitle{$\alpha$ Cen B b Mass}
\shortauthors{PLAVCHAN, CHEN, \& POHL}

\begin{document}

\title{What is the mass of $\alpha$ Cen B \lowercase{b}?}

\author{Peter Plavchan\altaffilmark{1}, Xi Chen\altaffilmark{2}, and Garrett Pohl\altaffilmark{1}}

\altaffiltext{1}{corresponding author; peterplavchan@missouristate.edu, Department of Physics, Astronomy and Materials Science, 901 S National Ave, Springfield, MO 65897}
\altaffiltext{2}{NASA Exoplanet Science Institute, California Institute of Technology, M/C 100-22, 770 South Wilson Avenue, Pasadena, CA 91125}

\begin{abstract}
We investigate the possibility of constraining the sin $i$ degeneracy of $\alpha$ Cen B b -- with orbital period P=3.24 d; a = 0.042 AU; $m$ sin $i$ = 1.1 M$_\oplus$ -- to estimate the true mass of the newly reported terrestrial exoplanet in the nearest stellar system to our Sun.  We present detailed numerical simulations of the dynamical stability of the exoplanet in the $\alpha$ Cen AB binary system for a range of initial inclinations, eccentricities, and semi-major axes.  The system represents a benchmark case for the interplay of the Kozai mechanism, general relativistic and tidal forces.  From our simulations, there is only a small boundary in initial inclinations and initial semi-major axes that result in the migration via the Kozai mechanism of $\alpha$ Cen B b to its present location.  Inside this boundary, the planet orbit is stable for up to 1 Gyr against the Kozai mechanism, and outside this boundary the planet collides with $\alpha$ Cen B or is ejected.  In our three simulations where the planet migrates in towards the star via the Kozai mechanism, the final inclination is 46$^\circ$--53$^\circ$ relative to the AB orbital plane, lower than the initial inclination of 75$^\circ$ in each case. We discuss inclination constraints from the formation of $\alpha$ Cen B b in situ at its present location, migration in a proto-planetary disk, or migration in resonance with additional planets.  We conclude that $\alpha$ Cen B b probably has a mass of less than 2.7 $M_\oplus$, implying a likely terrestrial composition warranting future confirmation. 
\end{abstract}

\keywords{planetary systems: formation --- planetary systems: dynamical evolution and stability ---  planets and satellites: individual: $\alpha$ Centauri}

\section{Introduction}

Over 1800 exoplanets have been discovered over the past 25 years to orbit stars other than our Sun (NASA Exoplanet Archive, Akeson et al. 2013), a remarkable achievement enabled by continued advances in precise instrumentation and calibration, cadence and observational strategies, and computational analysis techniques.  The discovery of a 1.13 M$_\oplus$sin$i$ terrestrial planet in a 3.2 day orbit around the K1V dwarf $\alpha$ Cen B by Dumusque et al. 2012 (hereafter D12) is an exemplary case, with a reported velocity semi-amplitude of 51 cm/s and a reported uncertainty of only 4 cm/s.  Substantial care is taken by D12 in the characterization of stellar activity that dominates the radial velocity signal, and in the understanding of instrumental systematic errors, to take advantage of binning high cadence observations.  This discovery not only represents the closest known exoplanet to our Sun, but also the lowest mass planet with the smallest Doppler signature detected with the radial velocity method to date.  The masses of terrestrial exoplanets in multiple exoplanet systems have also been measured around more distant stars via transit timing variations (e.g., Steffen et al. 2012, Marcy et al. 2014).  Much work remains to be done in confirming this detection, and in this paper we assume the detection is robust.

One of the unavoidable limitations of the radial velocity method is the sin$i$ inclination degeneracy in the mass of the exoplanet that results from observing only the velocity component of the stellar reflex motion that is projected along our line of sight.  It is critical to resolve this inclination degeneracy to directly constrain the mass of $\alpha$ Cen B b and to confirm that it is definitively a planet just slightly more massive than the Earth.  The most direct approach to determine the inclination of the $\alpha$ Cen B b orbit with respect to our line of sight would be to confirm or rule out transits of the exoplanet in front of $\alpha$ Cen B.  However, these observations have not yet been published, and will be challenging even from space given the required precision, the expected transit duration, and the brightness of $\alpha$ Cen B.  Additionally, the radial velocity observations in D12 lack the precision necessary to detect the expected Rossiter-McLaughlin signature of a transiting planet (e.g., Winn et al 2010), although that does not detract from the significance of this discovery.  Another common approach to constrain the orbital inclination with respect to our line of sight is to invoke dynamical stability arguments for multiple exoplanet systems (Fang et al. 2012).  However, these arguments do not apply to this system with only one identified exoplanet to date.

In this work, we invoke dynamical modeling and observational arguments to constrain the inclination of the orbit of $\alpha$ Cen B b.  We first present our dynamical simulations and results.  Next, we review constraints on the inclination of  $\alpha$ Cen B b that can be inferred from the literature.  Finally, we suggest future work to constrain the inclination and consequently the true mass of $\alpha$ Cen B b.

\section{Numerical Integrations of the $\alpha$ Cen AB, $\alpha$ Cen B \lowercase{b} System}

$\alpha$ Cen is one of the most well-characterized stellar systems due to its proximity to the Sun.  The exoplanet host star is part of a $\sim$5 Gyr triple system.  The AB binary has an eccentricity of $\sim$0.52, a semi-major axis of $\sim$23.4 AU, an orbital period of ~80 years, a closest approach of $\sim$11.2 AU, and an inclination of 79.205 $\pm$ 0.041$^\circ$ on the sky so that it is viewed nearly edge-on (Yildez 2008, 2006, Morel et al. 2000, and references therein).  The less massive M dwarf C component Proxima Cen is at a relatively distant $\sim$15,000 AU (Wertheimer \& Laughlin 2006).     

We carry out 567 N-body simulations of the $\alpha$ Cen AB system with a range of initial inclinations, eccentricities and semi-major axes for $\alpha$ Cen B b.   The simulations are carried out in the rest frame of $\alpha$ Cen B.   We start with the \textit{Mercury6} N-body integrator code (Chambers 1999, Chambers \& Wetherill 1998).   The original \textit{Mercury6} code does not include corrections for the tidal circularization and general relativistic precession of bodies close to the host star, as is relevant for $\alpha$ Cen B b.  Thus, we added these functions to the \textit{Mercury6} Fortran code in stubbed place-holder functions while using the \textit{bs} and \textit{radau} integrators.   

For the tidal interactions between the host star and exoplanet, we implement Equations 10 and 11 from Rodriguez et al. (2011), taken in the limit that the planet mass is much less than the host star mass, accounting for star-planet and planet-star tides (e.g. ignoring planet-planet tides for multi-planet systems).  We assume a tidal Q of 100 for $\alpha$ Cen B b and 10$^6$ for $\alpha$ Cen B, as is typically estimated for terrestrial planets and stars respectively (Bodenheimer et al. 2003, Wu 2003, Penev et al. 2012, Pena 2010, Ray et al. 1996, Lagus \& Anderson 1968, Goldreich \& Nicholson 1977, Goldreich \& Soter 1966).  An error in the planet Q value will translate into an error in the tidal circularization time-scale, and we verify our implementation by recovering the tidal circularization timescale for the Jovian exoplanet HD 209458b of $\sim$82 Myr from Bodenheimer et al.(2003).

For general relativistic corrections, we implement the correction based upon Rodriguez et al. (2011) and Buetler (2005), ignoring the precession of the central body.  We verify the accuracy of our implementation with the known precession of Mercury in the Solar System.  We also tried but did not use corrections from Danby (1962) which does not include a tangential component, and one from Vitagliano (1997).

We adopt the parameters in D12 for the period and planet mass, and AB stellar parameters (Figure 1).  We assume the effect of Proxima Cen to be negligible as is done in other analyses of the dynamical evolution of the $\alpha$ Cen system (Quintana et al. 2002).   We do not increase the mass of $\alpha$ Cen B b for increasing inclinations w/r/t to our line of sight, except to note this would have the effect of decreasing the stability and thus our simulations represent a more conservative estimate.  Using the relation $R=2 R_* (\rho_*/\rho_{pl})^{\frac{1}{3}}$,  and assuming the density of the Sun and the density of Earth for $\alpha$ Cen B and B b respectively, we calculate a Roche radius for $\alpha$ Cen B of 3.15 R$_*$ (0.012 AU).  We assume that the planet is tidally destroyed if the semi-major axis evolves to within this orbital distance, and we also assume that the planet does not survive a stellar collision.

\begin{figure*}
\includegraphics[width=0.42\textwidth,clip=true,trim=3cm 8cm 9cm 9cm]{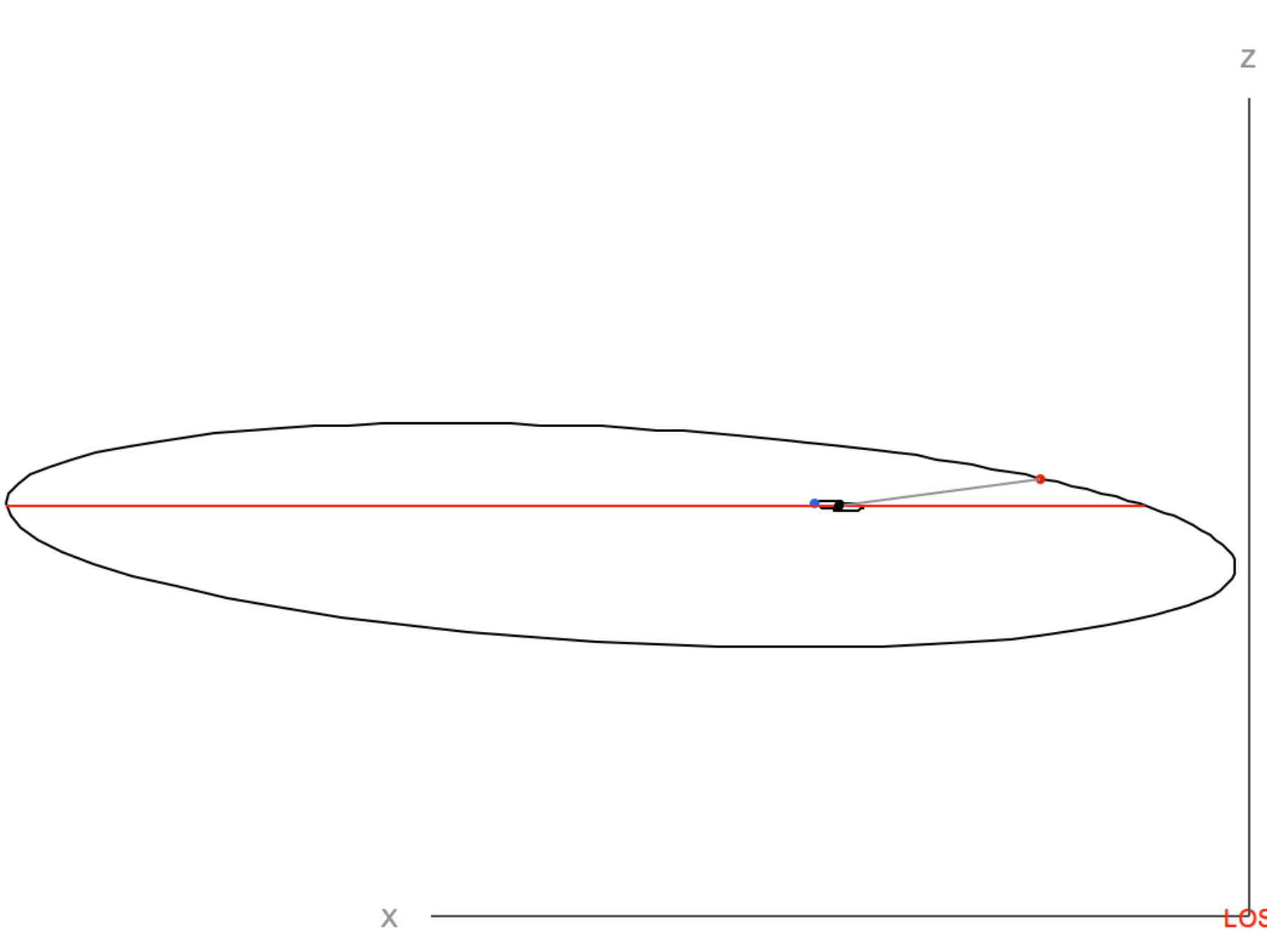}
\includegraphics[width=0.42\textwidth,clip=true,trim=3cm 8cm 9cm 9cm]{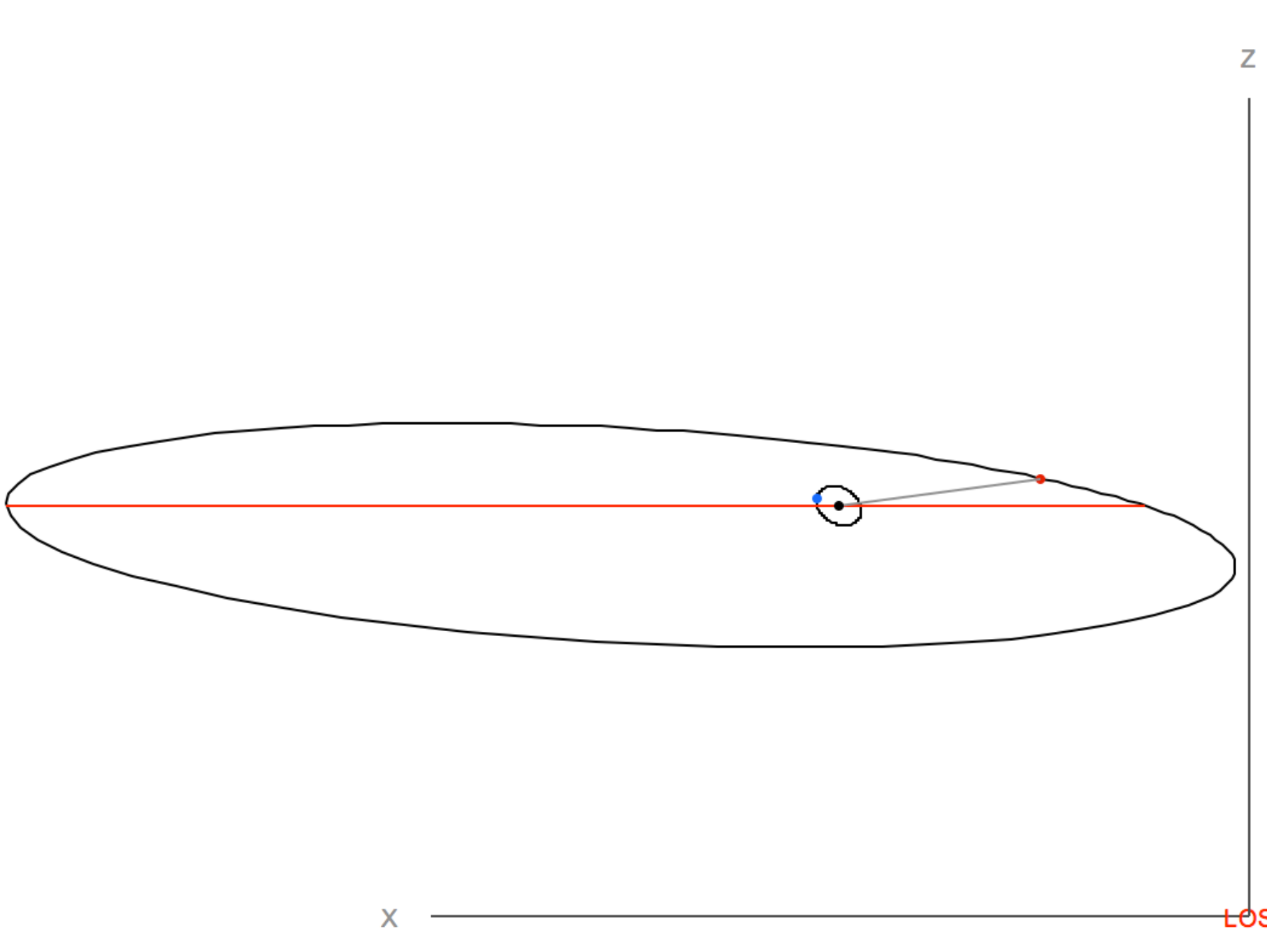}\\
\includegraphics[width=0.42\textwidth,clip=true,trim=8.5cm 8cm 3cm 3cm]{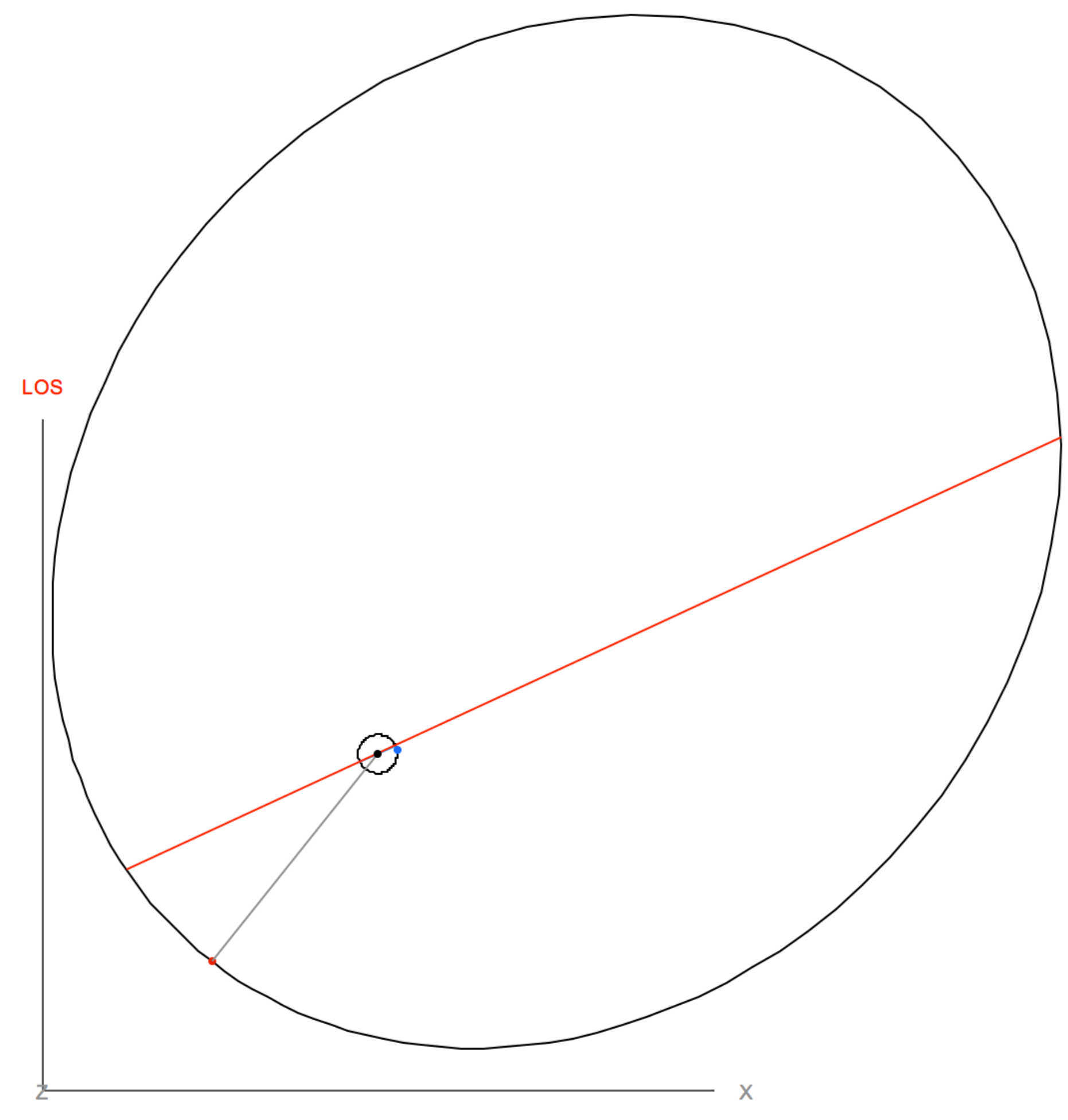}
\includegraphics[width=0.42\textwidth,clip=true,trim=8.5cm 8cm 3cm 3cm]{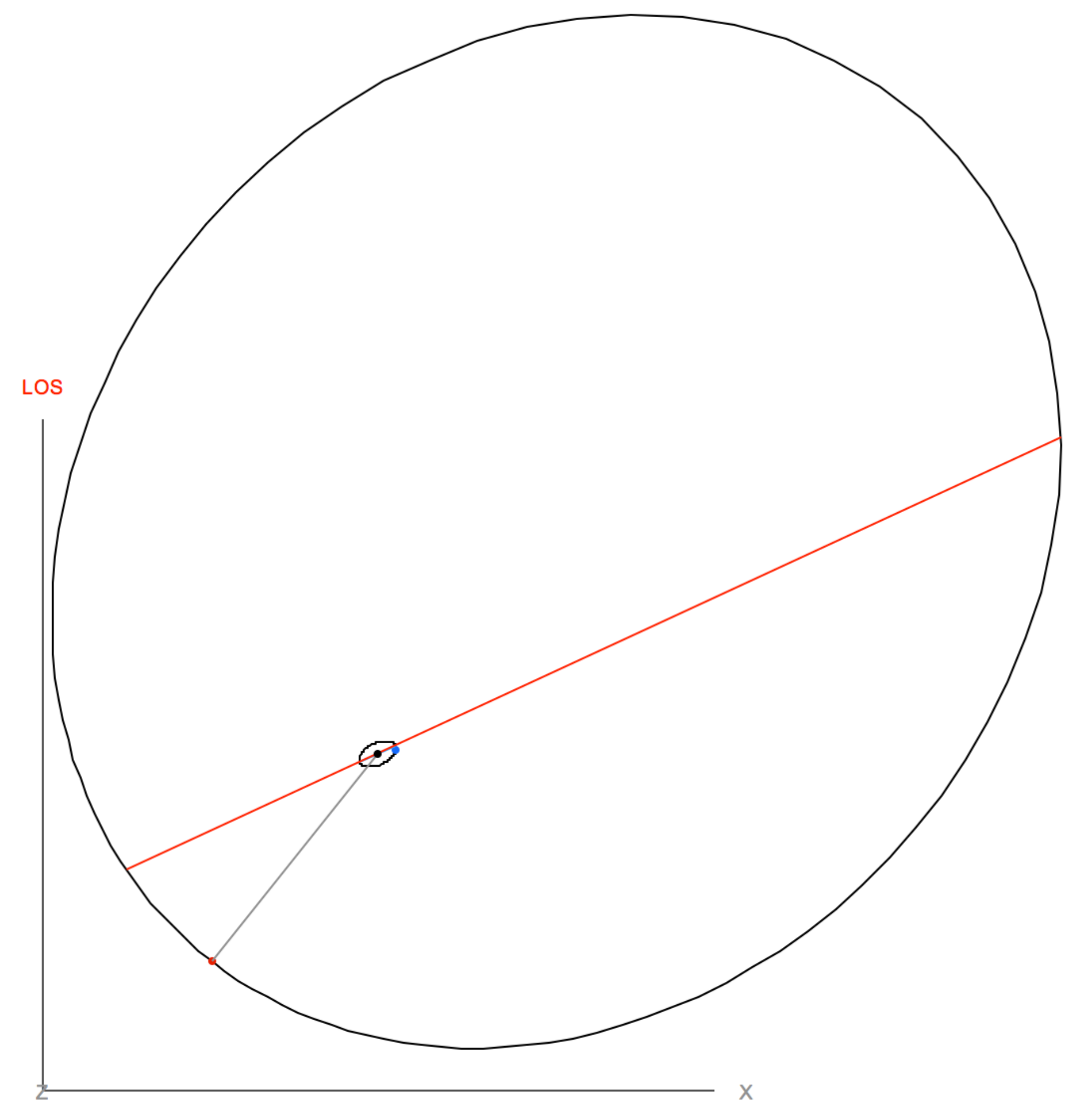} \\
\includegraphics[width=0.42\textwidth,clip=true,trim=6cm 3cm 6cm 5cm]{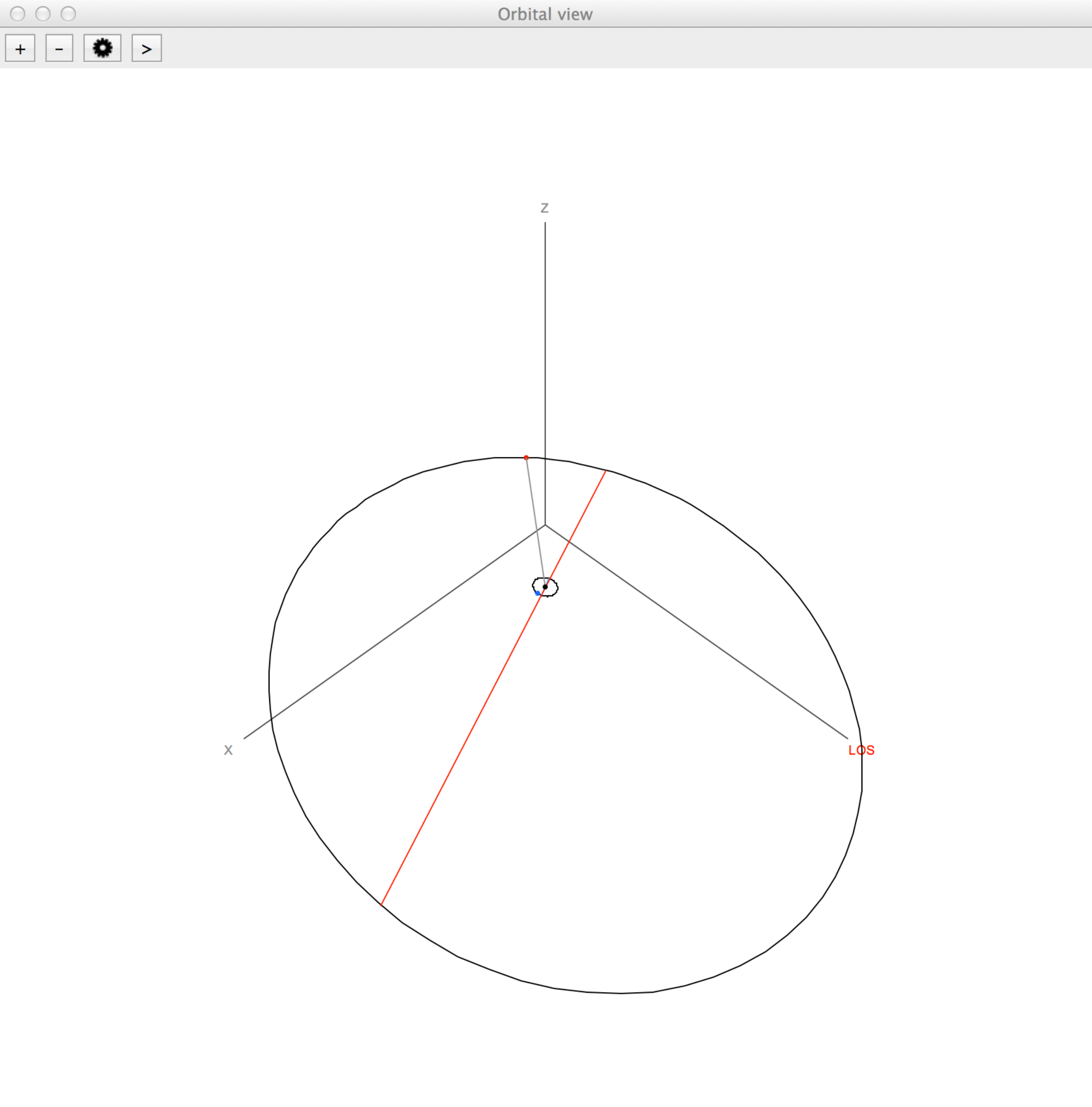}
\includegraphics[width=0.42\textwidth,clip=true,trim=6cm 3cm 6cm 5cm]{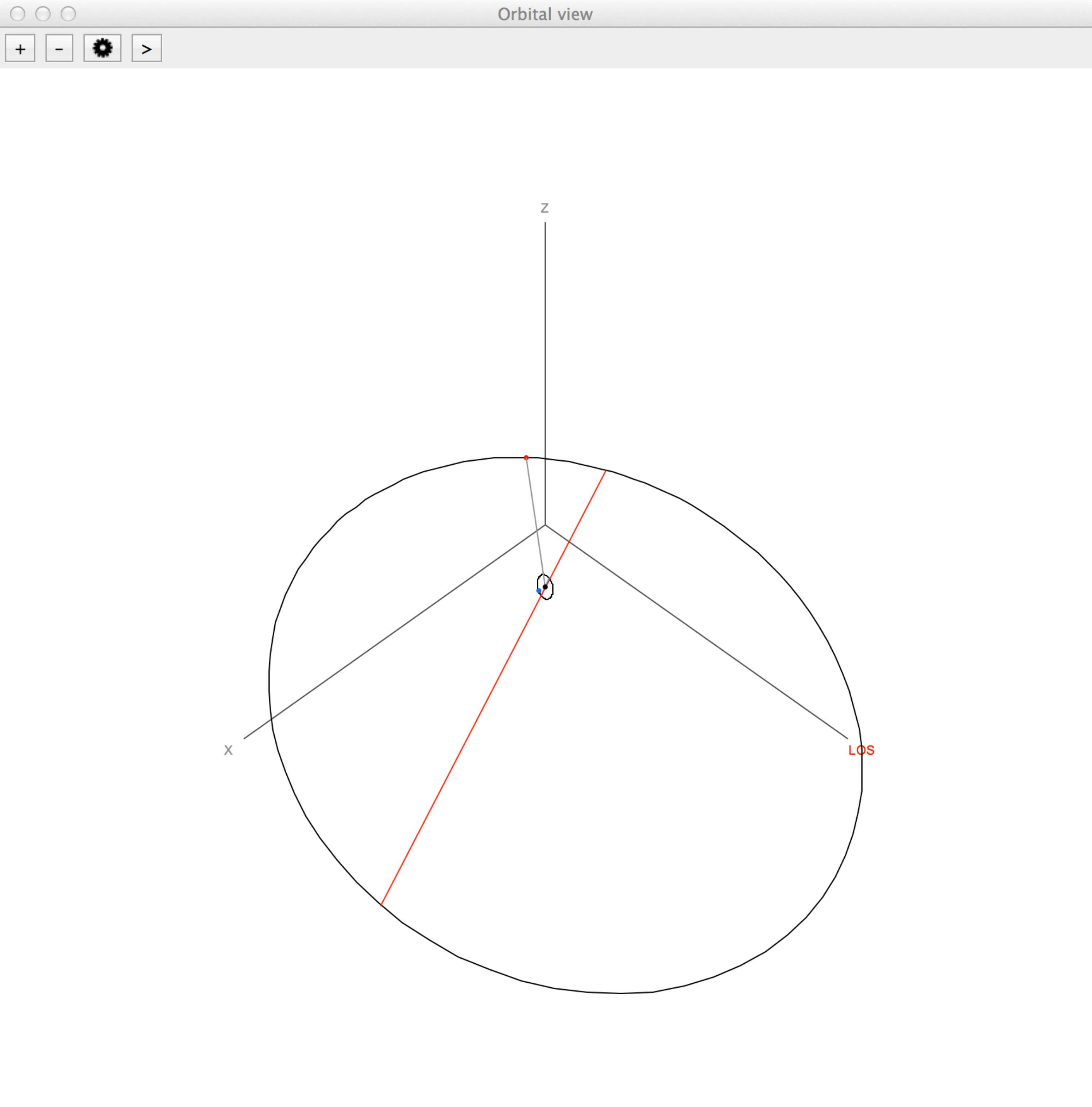}
\caption{Top to bottom: Line of sight (with an arbitrary rotation angle in the plane of the sky), top-down and isometric views of the $\alpha$ Cen A,B and Bb component orbits in the frame of reference of $\alpha$ Cen B.  The orbit for $\alpha$ Cen B b is enlarged by a factor of 20 in semi-major axis to help visualize the scale of the orbit.  The left side plots show an orbit for $\alpha$ Cen B b aligned with the AB orbital plane, and the right side plots show an orbit for $\alpha$ Cen B b inclined 45$^{\circ}$ w/r/t to the AB orbital plane, as carried out in our dynamical simulations.  The axis marked ``LOS'' in red indicates the line of sight to the Earth.  Plots produced with the Systemic Console (oklo.org; Meschiari et al. 2009).}
\end{figure*}

For 91 simulations, we place $\alpha$ Cen B b at its current semi-major axis of 0.042 AU, with relative inclinations to the AB orbital plane of 0, 15, 30, 45, 60, 75, 90, 105, 120, 135, 150, 165 and 180 degrees (e.g. both prograde and retrograde orbits are considered), and eccentricities of $e=$0, 0.1, 0.2, 0.3, 0.4, 0.5, and 0.6.  Orbits that achieve e$>$0.7 fall within the Roche radius of the $\alpha$ Cen B.  We ran these 91 simulations four times -- [1] with general relativistic precession and tidal forces, [2] with general relativistic precession but no tidal force, [3] with tidal force but no general relativistic precession, [4] without either general relativistic precession or tidal forces, for a total of 364 simulations.  Each of these 364 simulations were carried out for a duration of 2 Myr.

Next, we carry out 196 simulations by placing $\alpha$ Cen B b at different formation initial semi-major axes of 0.1, 0.2, 0.3, \dots, 1.8, 1.9, 2.0, 2.25, 2.5, 2.75, 3.0, 3.25, 3.5, 3.75 and 4 AU, with different initial inclinations of 0,15,30,45,60,75 and 90 degrees with respect to the AB orbital plane, and with initially circular orbits.  All simulations $\geq$0.2 AU were carried out for a duration of 1 Gyr, and the simulations at 0.1 AU were halted after $\sim$250 Myr.  Finally, we carry out 7 simulations with a fourth fictional equal mass planet in a coplanar 2:1 orbital resonance with $\alpha$ Cen B b with inclinations of 0, 15, 30, 45, 70 and 90 degrees with respect to the AB orbital plane.  Because the simulations involve short dynamical time-scales due to the 3.2 day period of $\alpha$ Cen B b, we ran our simulations on the NASA Exoplanet Science Institute ``bluedot'' 128-core cluster, using the local disks on each node to avoid network disk bottlenecks in the computation time.  Simulation orbital parameters are recorded in ASCII text to disk every 100 years, which result in $\sim$1 GB of data per simulation.

\section{Numerical Simulation Results}

We present our simulation results in this section.  Figures 2 \& 3 show representative simulations with $\alpha$ Cen B b at its present semi-major axis of 0.042 AU, with and without general relativistic and tidal force corrections.  Figures 4--7 show representative simulations with $\alpha$ Cen B b formed at different initial semi-major axes of 0.2, 0.5, 1 and 2 AU.  Figures 8-12 present the outcomes of all simulations.

We find that within 0.1 AU, the precession of $\alpha$ Cen B b's orbit due to general relativity dominates the dynamical evolution of the system, leading to stable orbits at all orbital inclinations relative to the orbital plane of the $\alpha$ Cen AB binary for up to 250 Myr.  Without GR precession and tidal forces, the Kozai mechanism would eject planets with inclinations $>$60$^\circ$ at the current semi-major axis of $\alpha$ Cen B b.

For a simulated planet formation location of 0.2 AU, the Kozai mechanism significantly alters the orbit of $\alpha$ Cen B b for inclinations $>$60$^\circ$, resulting in migration, ejection or collision with $\alpha$ Cen B, even with GR precession and tidal forces included.  For simulated planet formation locations of 0.3--1.2 AU and relative inclination of $>$60$^\circ$, the Kozai mechanism significantly alters the orbit of $\alpha$ Cen B b.  We find the same outcome for simulated planet formation locations of 1.3--2.0 AU and relative inclinations of $>$45$^\circ$.  For simulated planets above the critical Kozai angle of 39.2$^\circ$ at an initial inclination of 45$^\circ$, and interior to 2.0 AU, the Kozai mechanism excites the eccentricity of the planet orbit without inducing migration.  For simulated planet formation locations of 2.25 and 2.5 AU, only planets with inclinations less than the critical Kozai angle of 39.2$^\circ$ survive, and external to 2.5 AU, no simulated planets survive at any inclination for more than a few Myr.     

Only three simulations resulted in a stable migration of $\alpha$ Cen B b to a circular orbit at a smaller semi-major axis.   The first starts off at a semi-major axis 0.2 AU and a relative inclination of 75$^\circ$, and the planet migrates to 0.035 AU (Figure 4, bottom right).   For the two simulations shown in Figures 13 and 14, the planet migrates to just exterior to the Roche radius of $\alpha$ Cen B in $\sim$10$^5$ years.  This rapid migration is potentially not survivable by the planet.  The change in orbital energy is $\sim$80 times the binding energy of an Earth mass and density planet, and this energy must be dissipated in the planet.  For the simulated planet at an initial inclination of 60$^\circ$ and semi-major axis of 1.3 AU shown in Figure 15, after $\sim$120 Myr the Kozai mechanism does induce a steady migration inward to $\sim$0.5 AU.  However, at that point the planet collides with $\alpha$ Cen B b rather than continuing its inward migration, as the eccentricity is not damped fast enough by the tidal circularization.  

Finally, for the seven simulations of $\alpha$ Cen B b in a 2:1 orbital resonance with a second planet of comparable mass, the planets orbital inclination is stable at all inclinations.  None of our simulations result in the improbable capture of $\alpha$ Cen B b by $\alpha$ Cen A.  This would have been an intriguing outcome to show that it was possible for $\alpha$ Cen B b to conversely form around $\alpha$ Cen A and get captured in its present orbit by $\alpha$ Cen B.

\begin{figure*}
\includegraphics[width=0.5\textwidth,clip=true,trim=0cm 0cm 0cm 0cm]{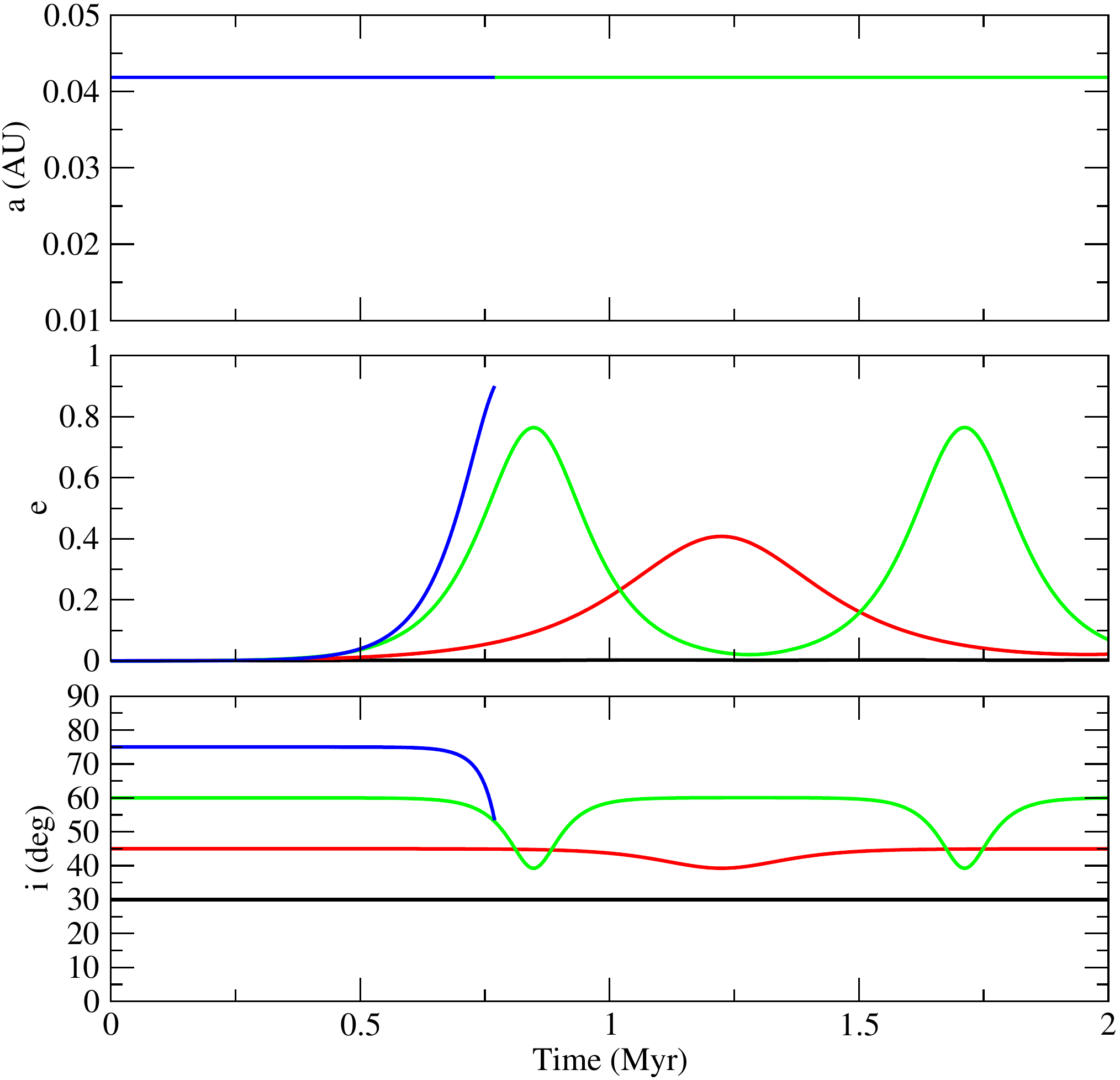}
\includegraphics[width=0.5\textwidth,clip=true,trim=0cm 0cm 0cm 0cm]{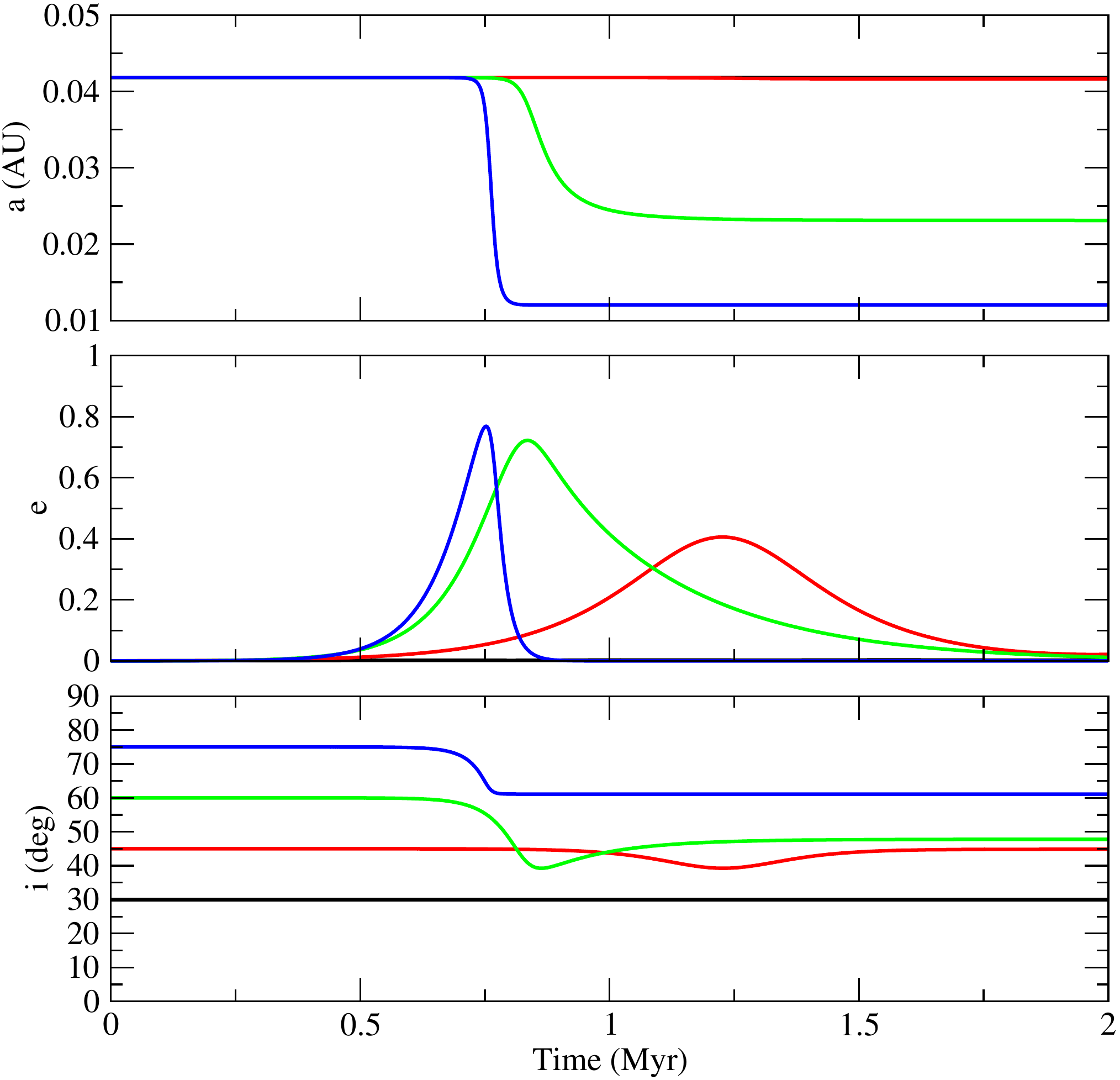}\\
\includegraphics[width=0.5\textwidth,clip=true,trim=0cm 0cm 0cm 0cm]{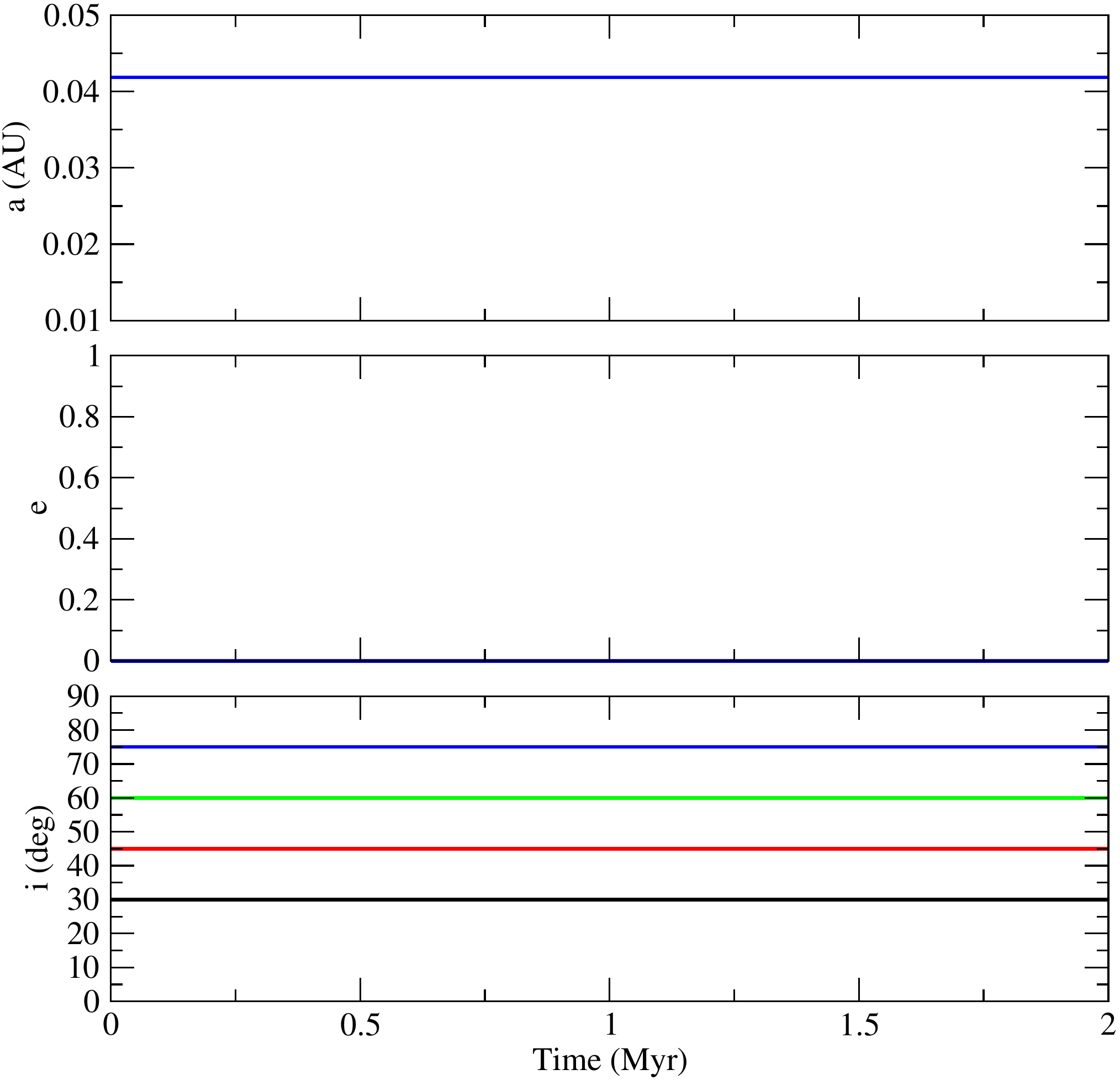}
\includegraphics[width=0.5\textwidth,clip=true,trim=0cm 0cm 0cm 0cm]{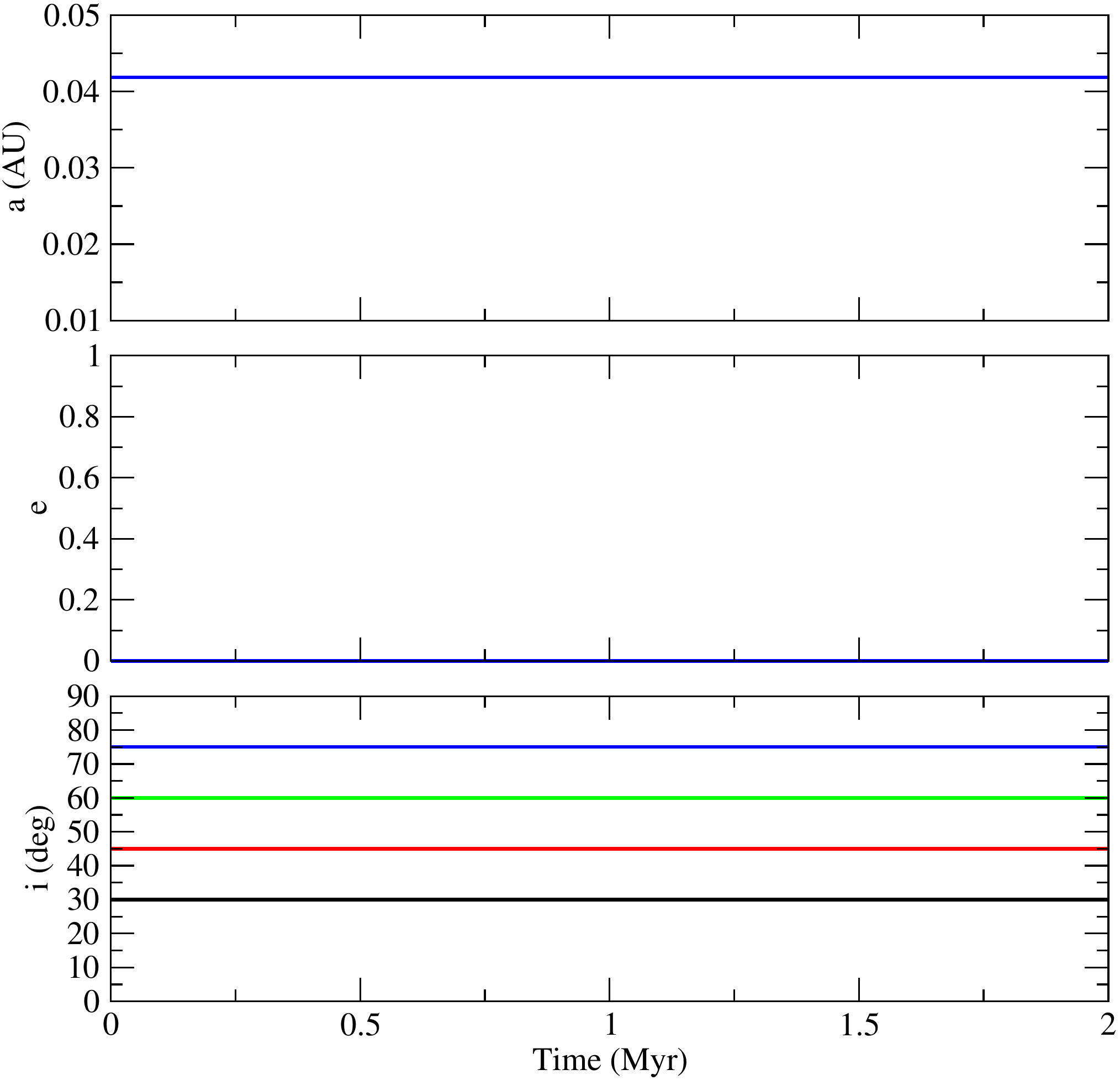}
\caption{Dynamical simulations for $\alpha$ Cen B b with an initial semi-major axis of 0.042 AU, an eccentricity of 0, and prograde inclinations of 30,45,60 and 75 degrees.  Other initial inclinations are not shown for clarity. The tri-panels from top left clockwise are the simulations [1] without general relativistic precession and without tidal forces, [2] without general relativistic precession and with tidal forces, [3] with both tidal forces and with general relativistic precession, [4] with general relativistic precession and without tidal forces, respectively.  In all plots, the different colors indicate different initial inclinations, and each simulation is carried out for a duration of 2 Myr.  Data points are plotted every 100 years and connected via line-segments.  Any simulation that terminates prior to 2 Myr resulted in the collision of the simulated $\alpha$ Cen B b with $\alpha$ Cen B or in some cases ejection.  These plots demonstrate the importance of general relativistic precession over the Kozai mechanism in the orbital evolution of $\alpha$ Cen B b at its present semi-major axis.}
\end{figure*}

\begin{figure*}
\includegraphics[width=0.5\textwidth,clip=true,trim=0cm 0cm 0cm 0cm]{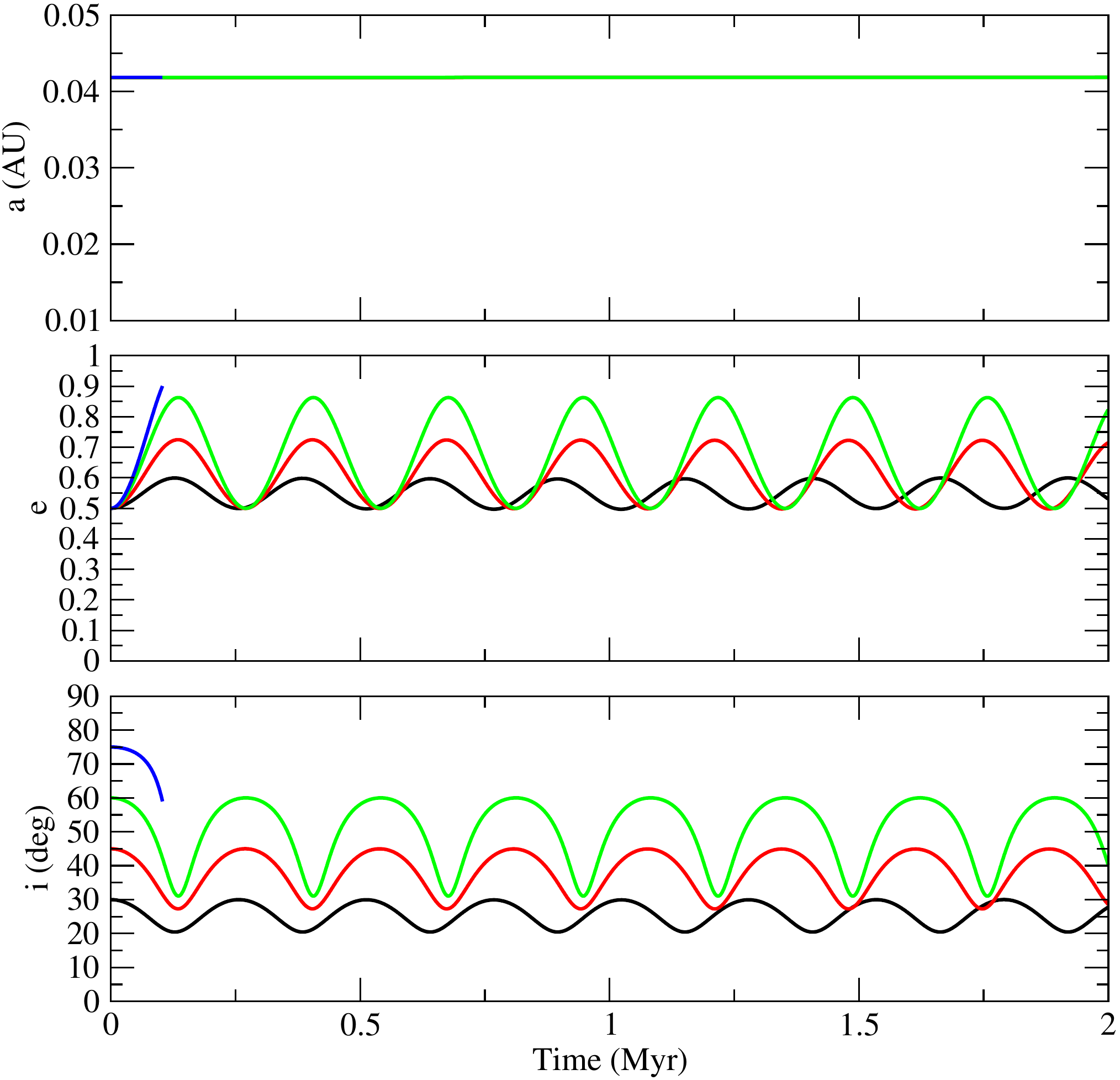}
\includegraphics[width=0.5\textwidth,clip=true,trim=0cm 0cm 0cm 0cm]{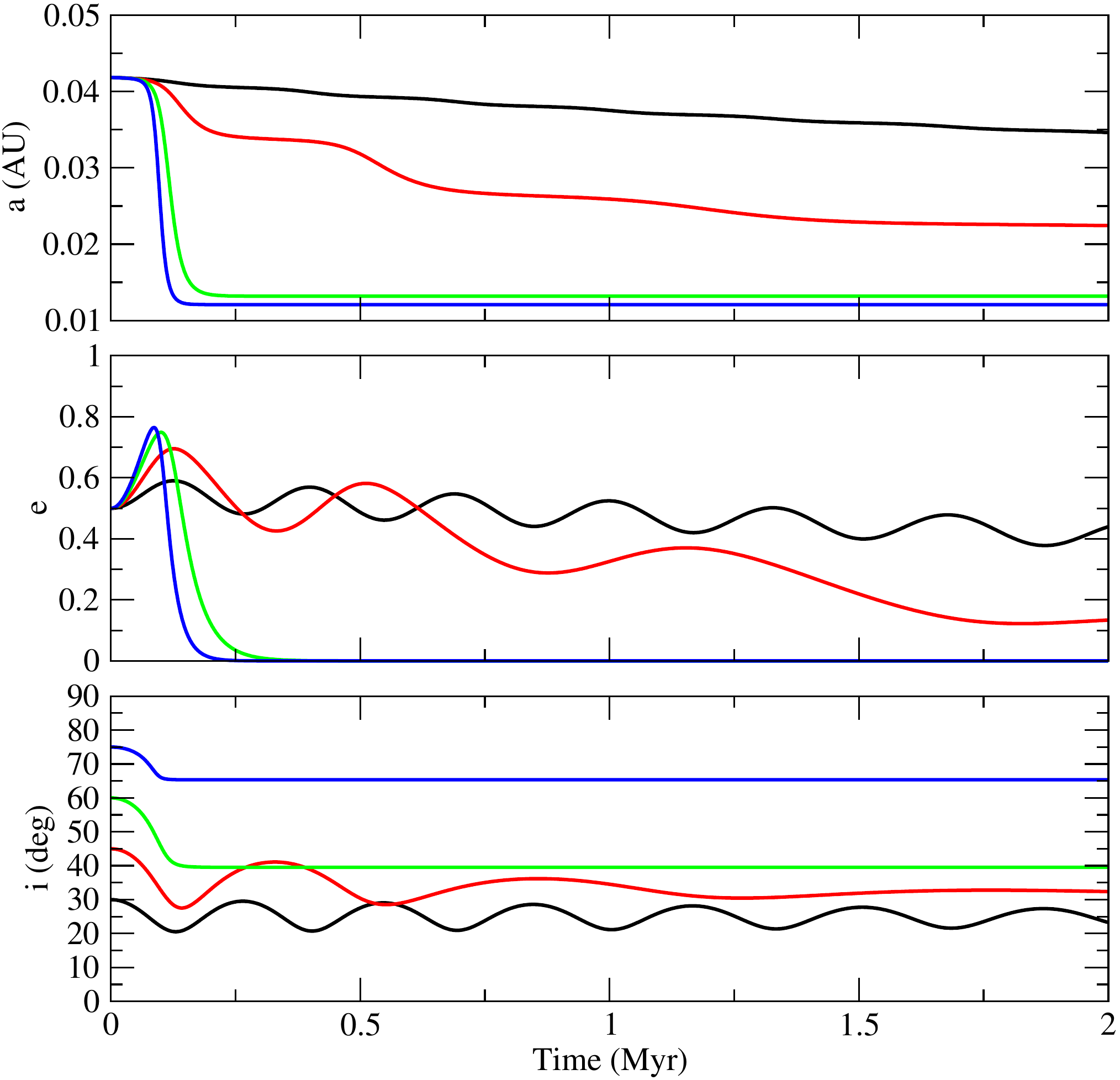}\\
\includegraphics[width=0.5\textwidth,clip=true,trim=0cm 0cm 0cm 0cm]{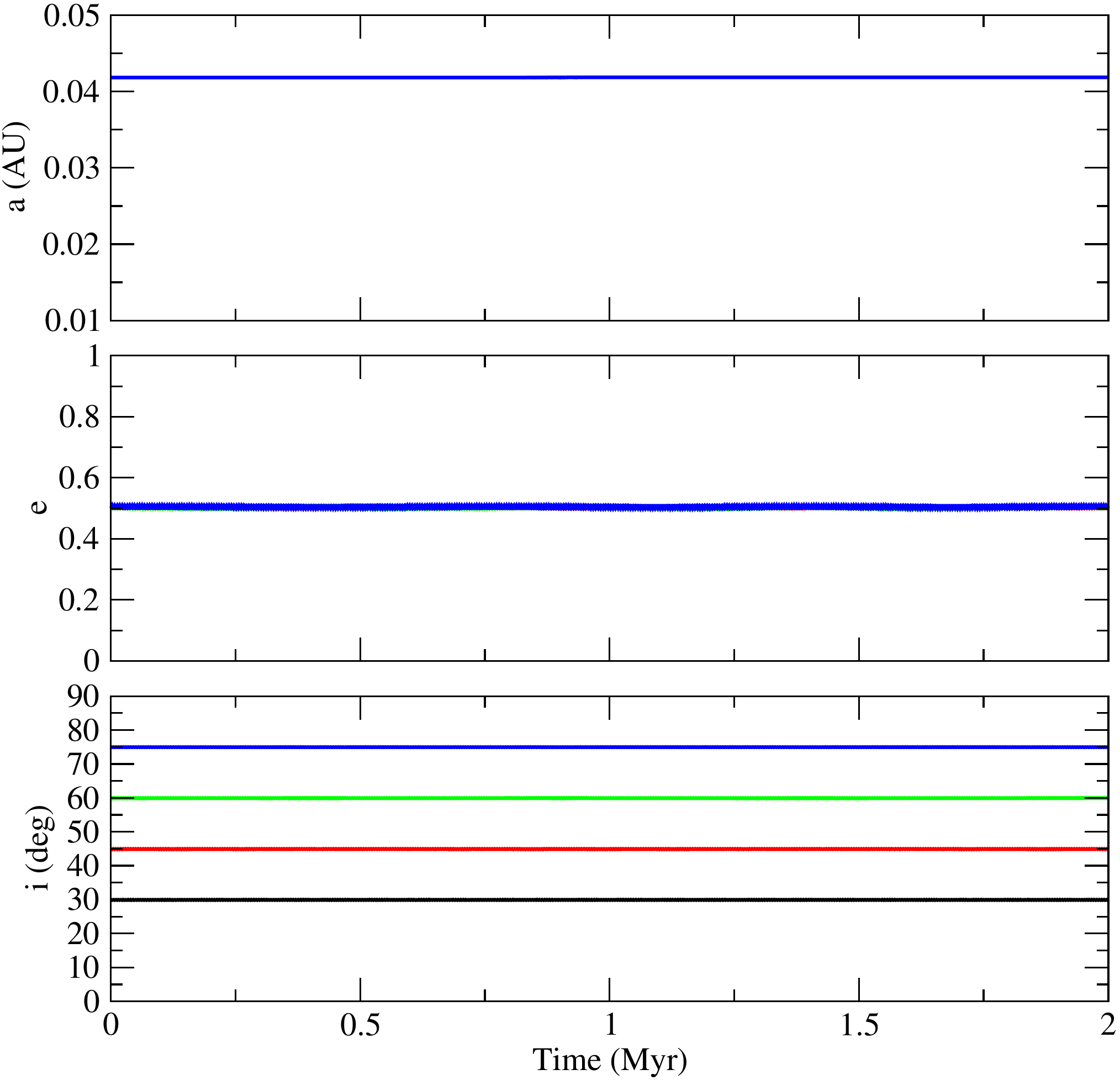}
\includegraphics[width=0.5\textwidth,clip=true,trim=0cm 0cm 0cm 0cm]{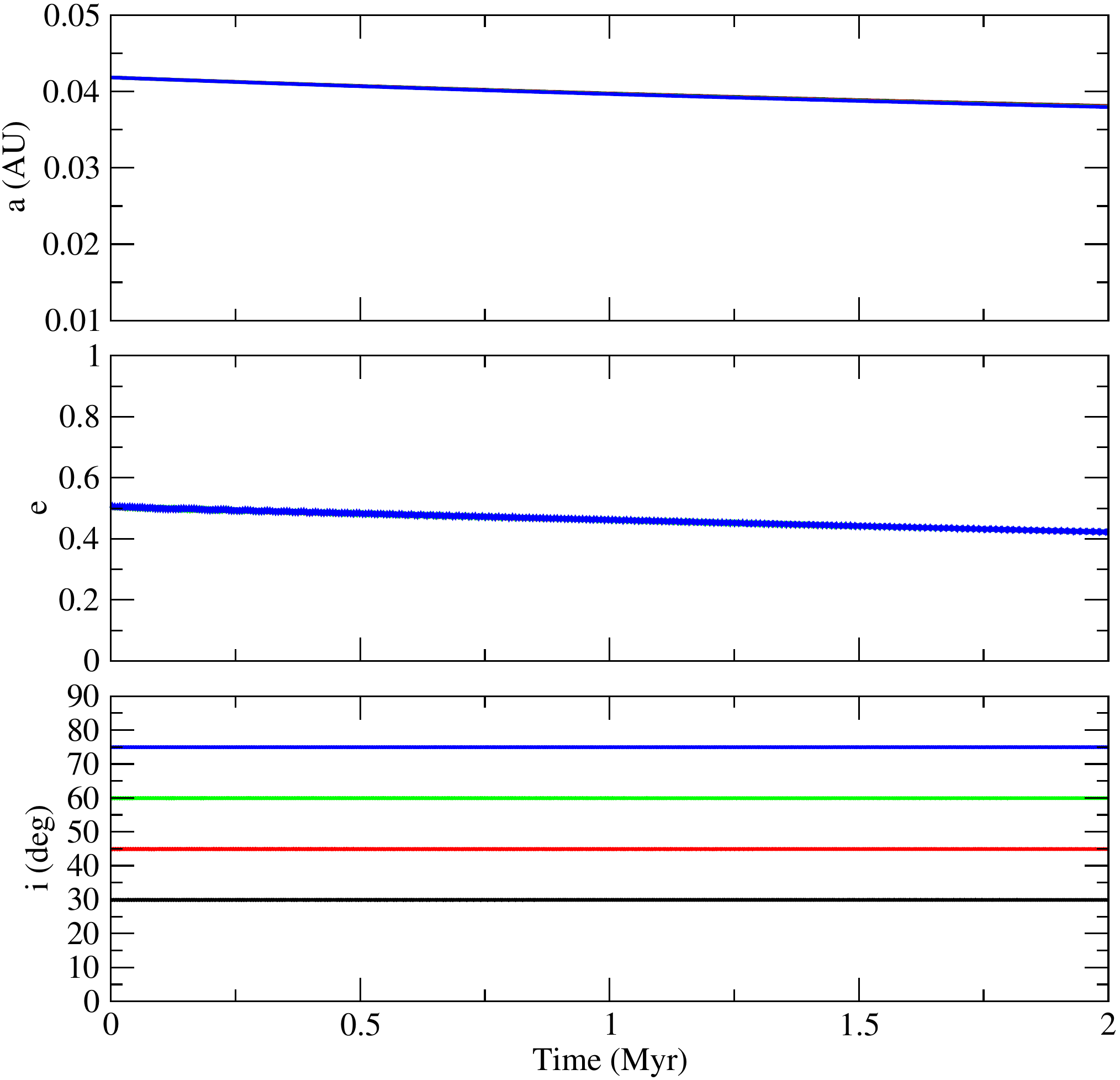}
\caption{The same as Figure 2, but with an initial eccentricity of 0.5 for all simulations plotted.}
\end{figure*}

\begin{figure*}
\includegraphics[width=0.5\textwidth,clip=true,trim=0cm 0cm 0cm 0cm]{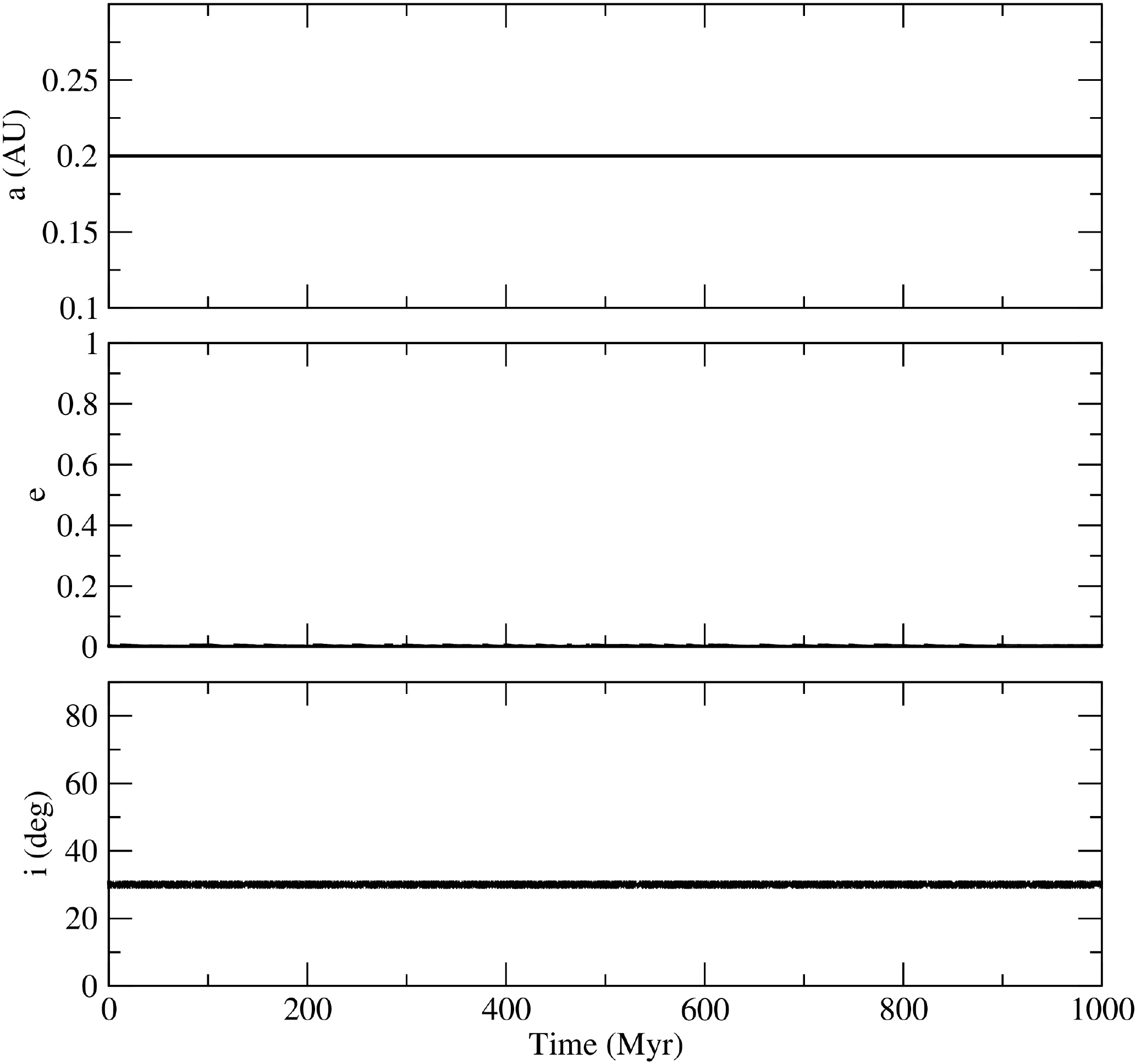}
\includegraphics[width=0.5\textwidth,clip=true,trim=0cm 0cm 0cm 0cm]{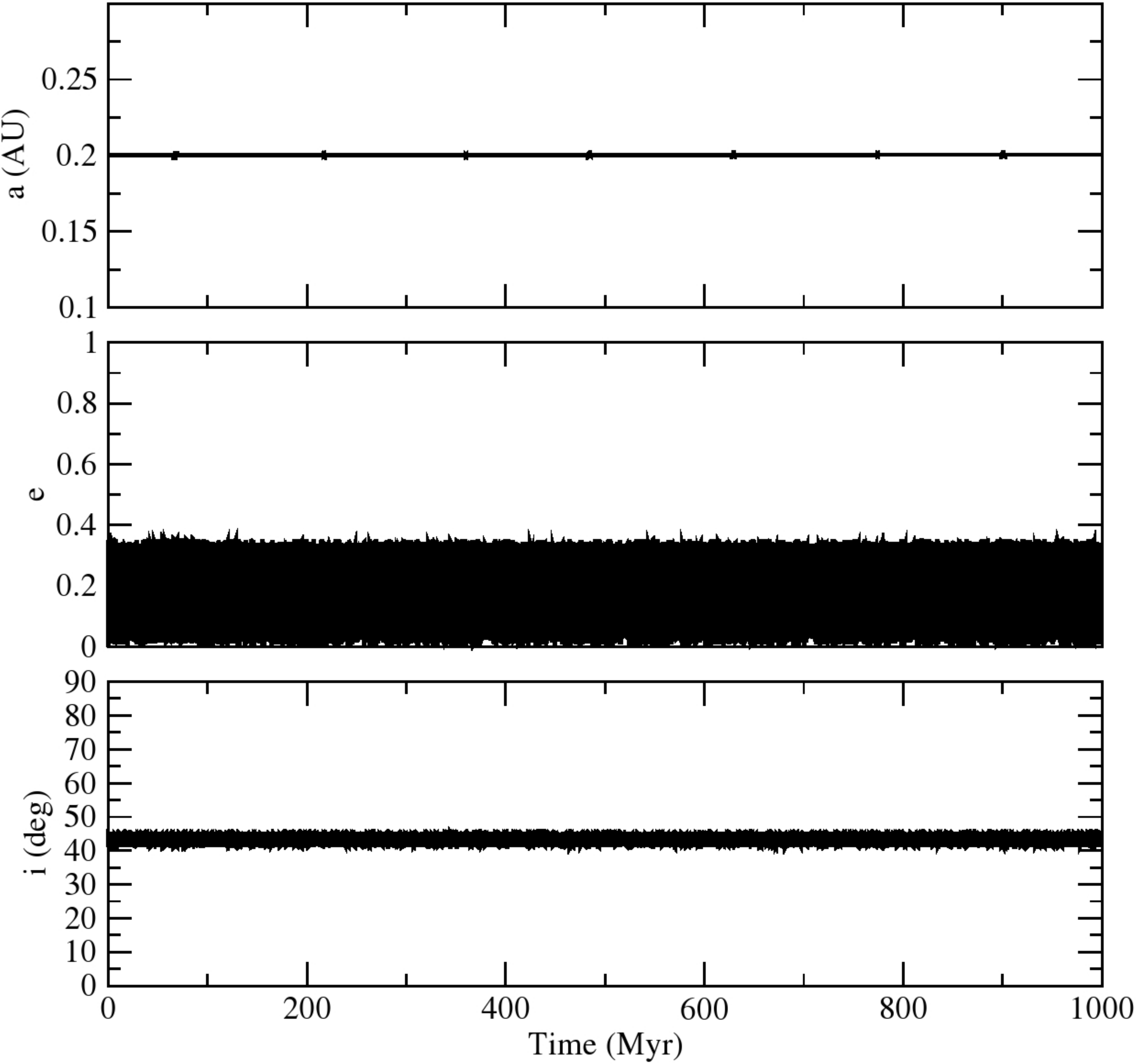}\\
\includegraphics[width=0.5\textwidth,clip=true,trim=0cm 0cm 0cm 0cm]{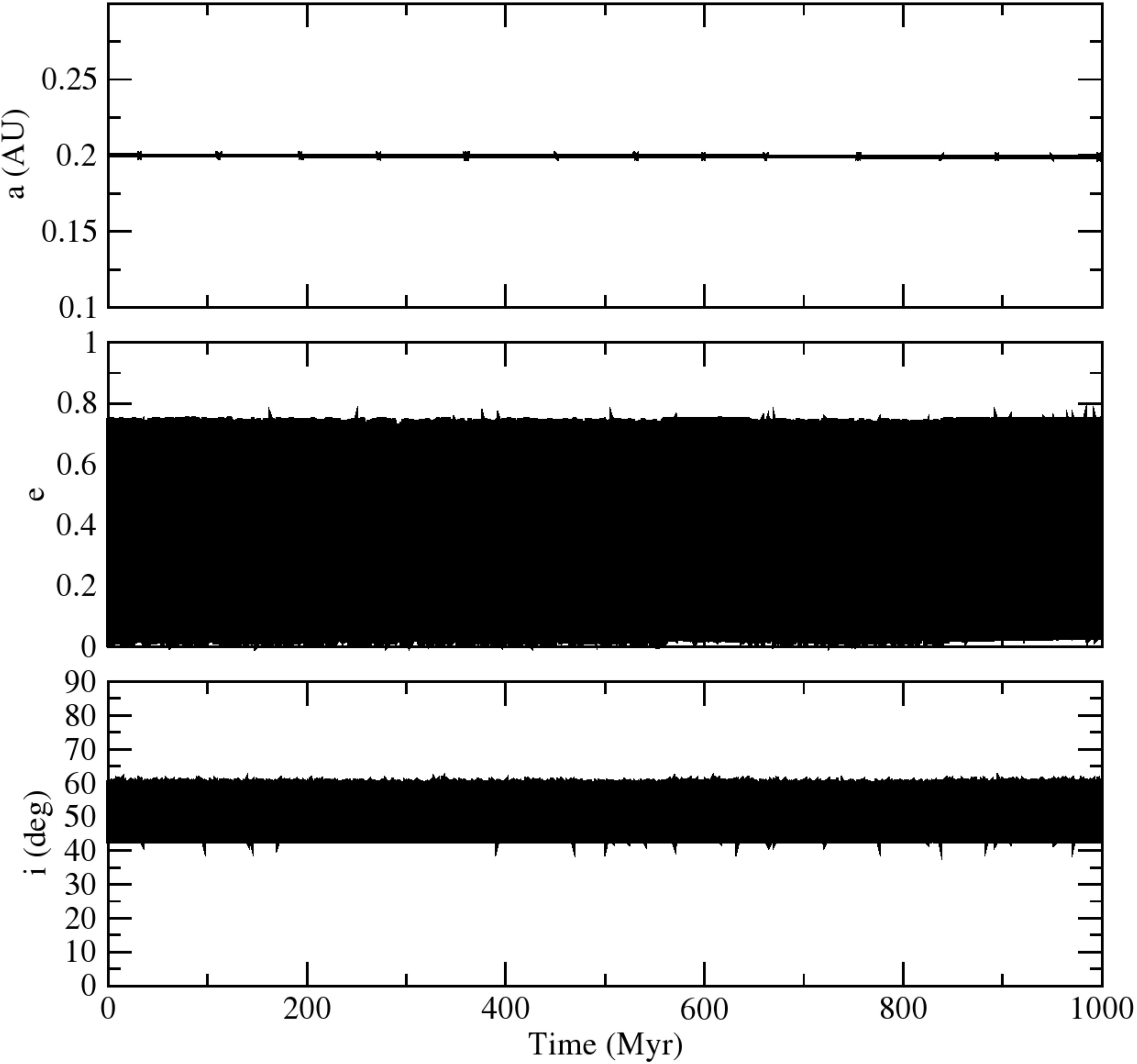}
\includegraphics[width=0.5\textwidth,clip=true,trim=0cm 0cm 5.5cm 1.5cm]{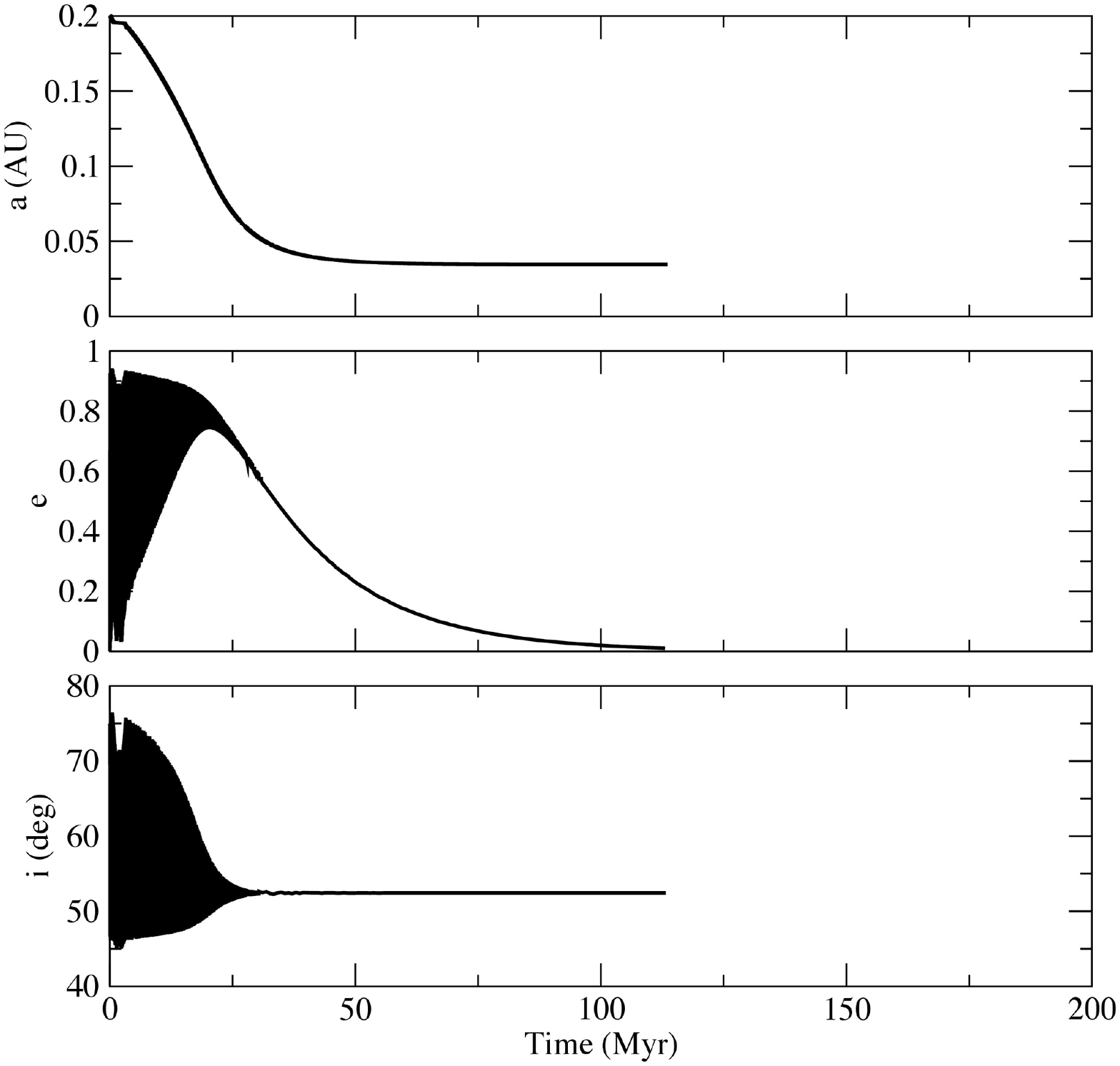}
\caption{Dynamical simulations for $\alpha$ Cen B b with an initial semi-major axis of 0.2 AU, an eccentricity of 0, and prograde inclinations of 30,45,60 and 75 degrees with respect to the AB orbital plane as indicated in the text.  The tri-panel plots shows the orbital evolution from top to bottom of the semi-major axis, eccentricity and inclination of $\alpha$ Cen B b including both tidal forces and general relativistic precession.  In all plots, each simulation is carried out for a duration of 1 Gyr.  Data points are plotted every 100 years and connected via line-segments.  Any simulation that terminates prior to 1 Gyr resulted in the collision of the simulated $\alpha$ Cen B b with $\alpha$ Cen B or ejection.}
\end{figure*}

\begin{figure*}
\includegraphics[width=0.5\textwidth,clip=true,trim=0cm 0cm 0cm 0cm]{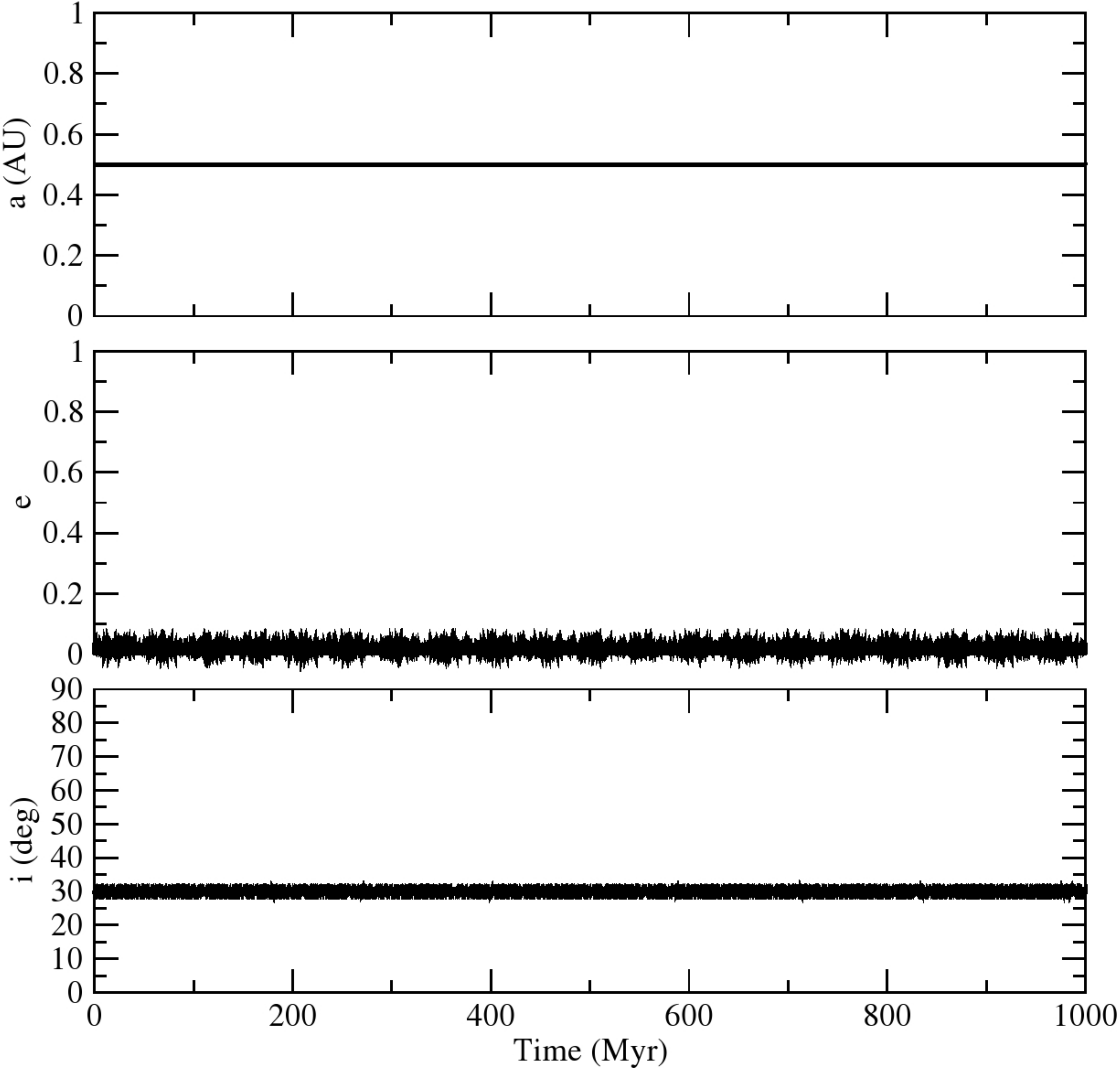}
\includegraphics[width=0.5\textwidth,clip=true,trim=0cm 0cm 0cm 0cm]{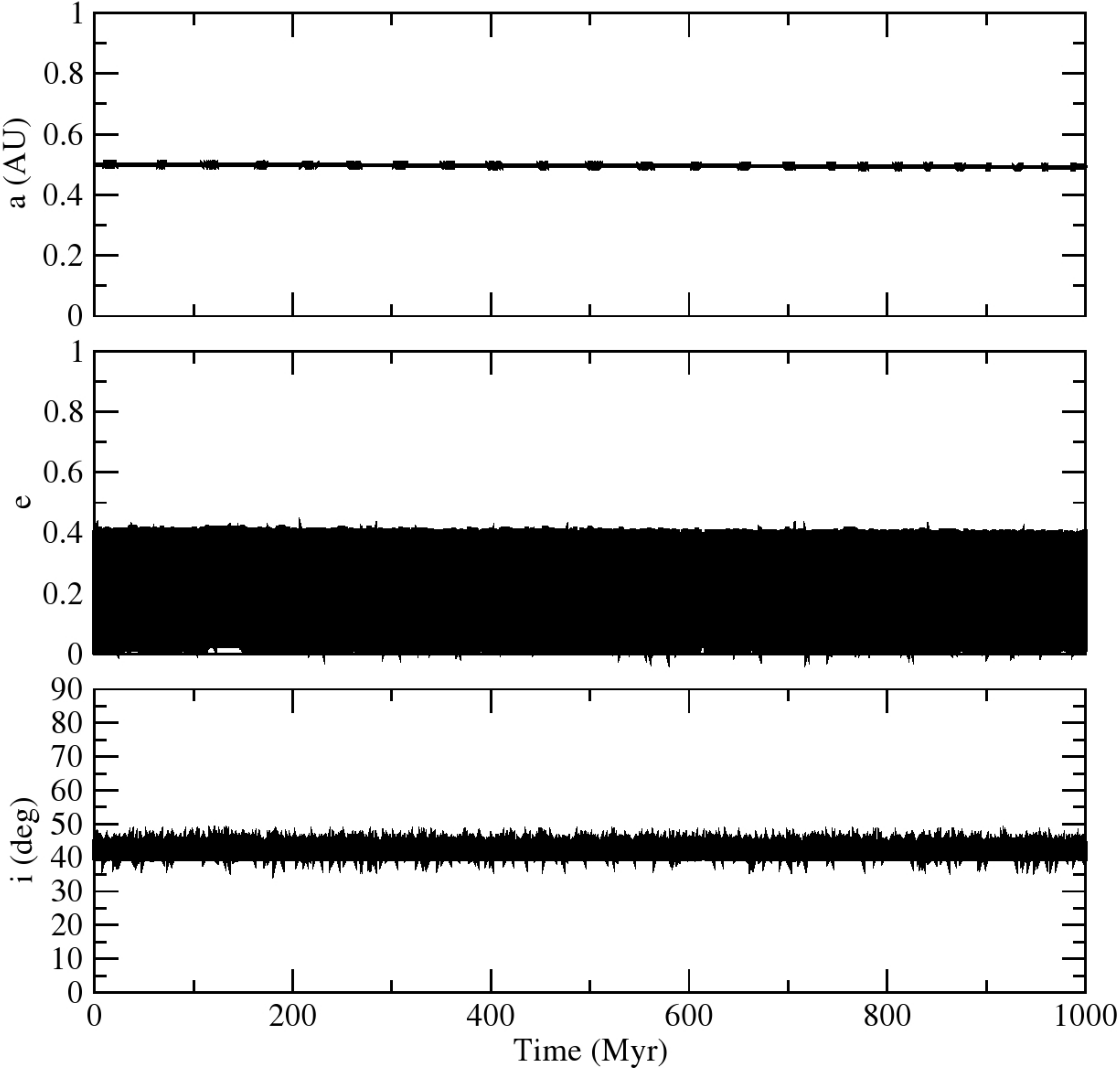}\\
\includegraphics[width=0.5\textwidth,clip=true,trim=0cm 0cm 0cm 0cm]{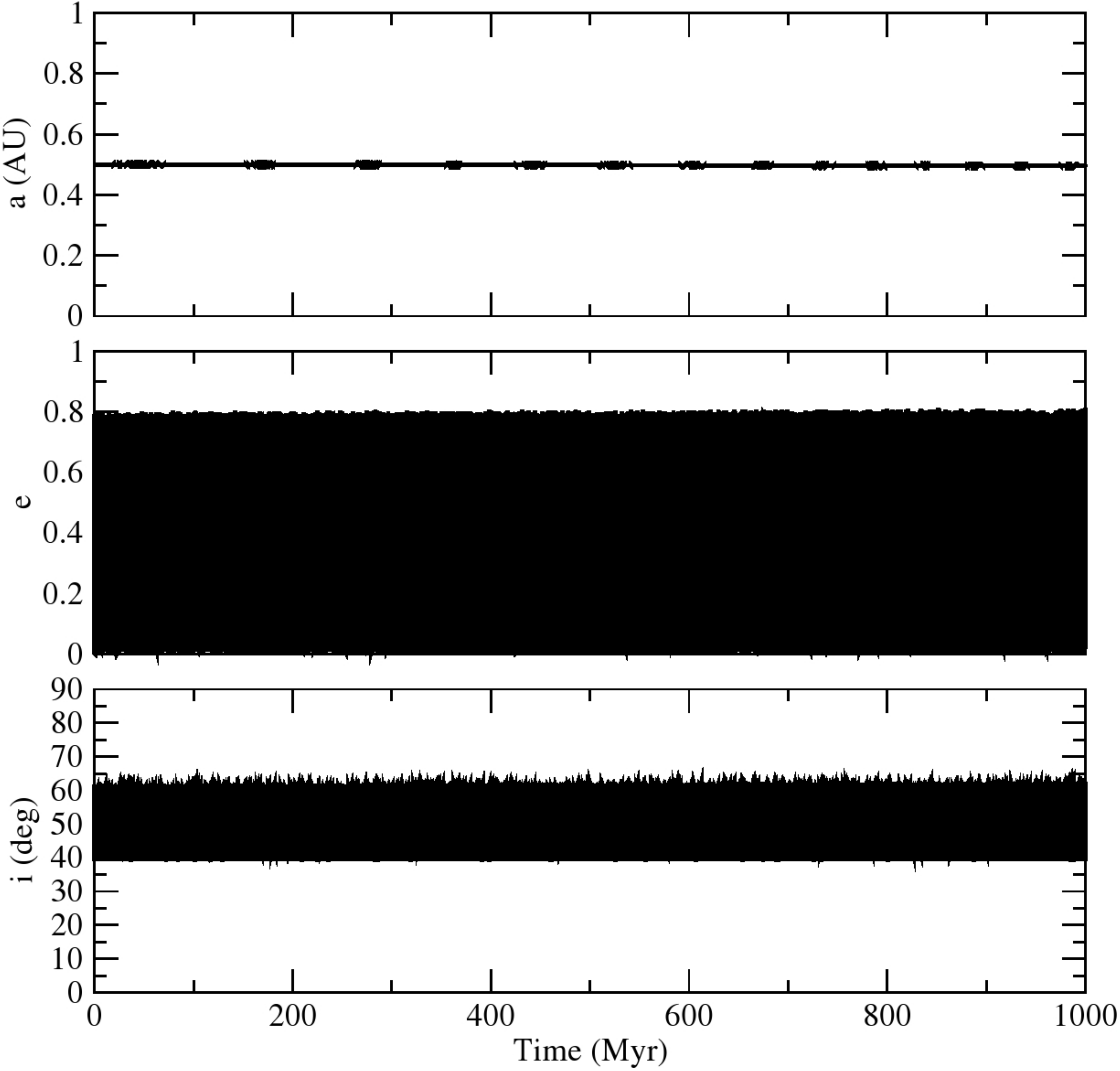}
\includegraphics[width=0.5\textwidth,clip=true,trim=0cm 0cm 0cm 0cm]{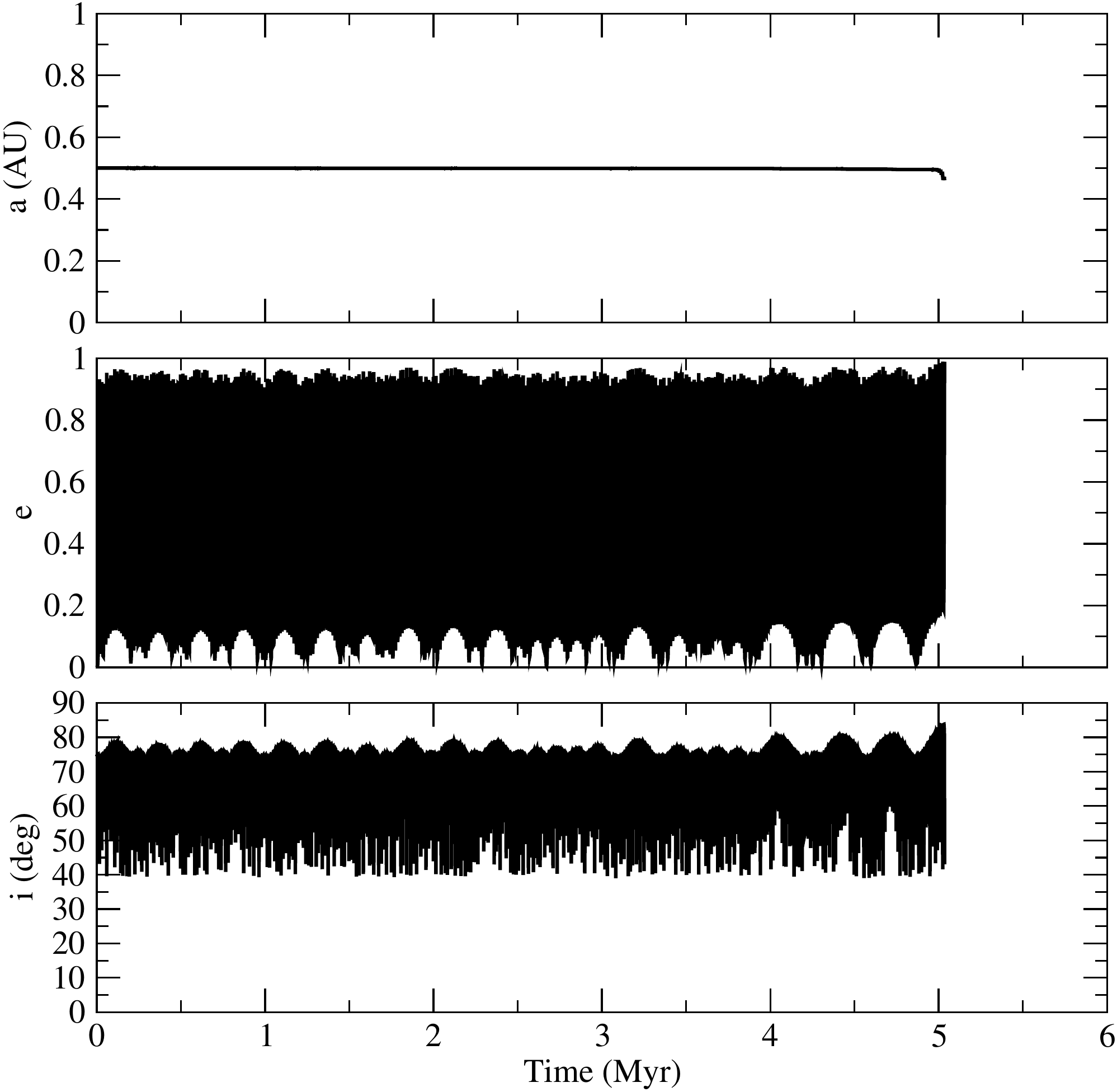}
\caption{The same as Figure 4, but with an initial semi-major axis of 0.5 AU.}
\end{figure*}

\begin{figure*}
\includegraphics[width=0.5\textwidth,clip=true,trim=0cm 0cm 0cm 0cm]{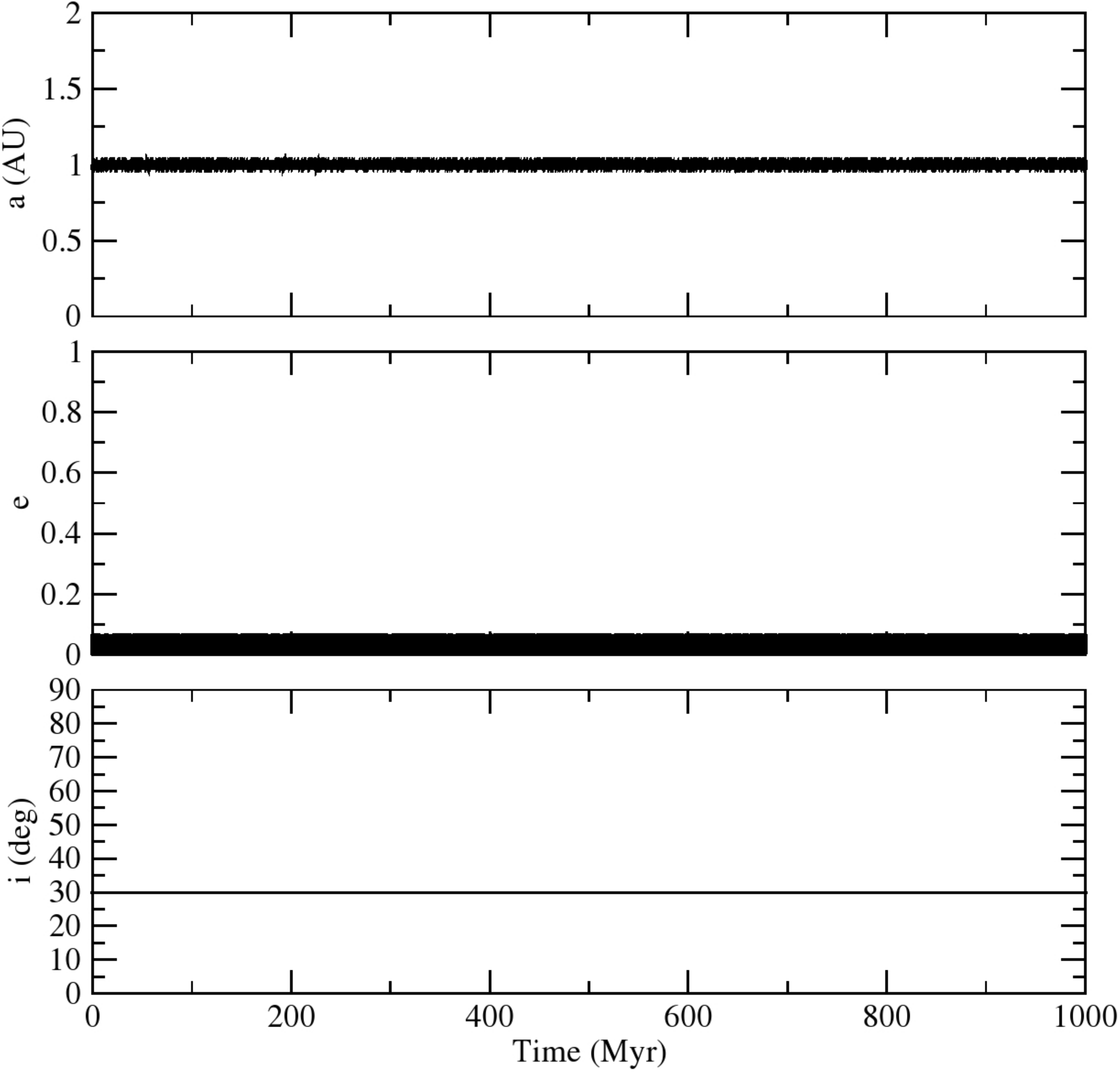}
\includegraphics[width=0.5\textwidth,clip=true,trim=0cm 0cm 0cm 0cm]{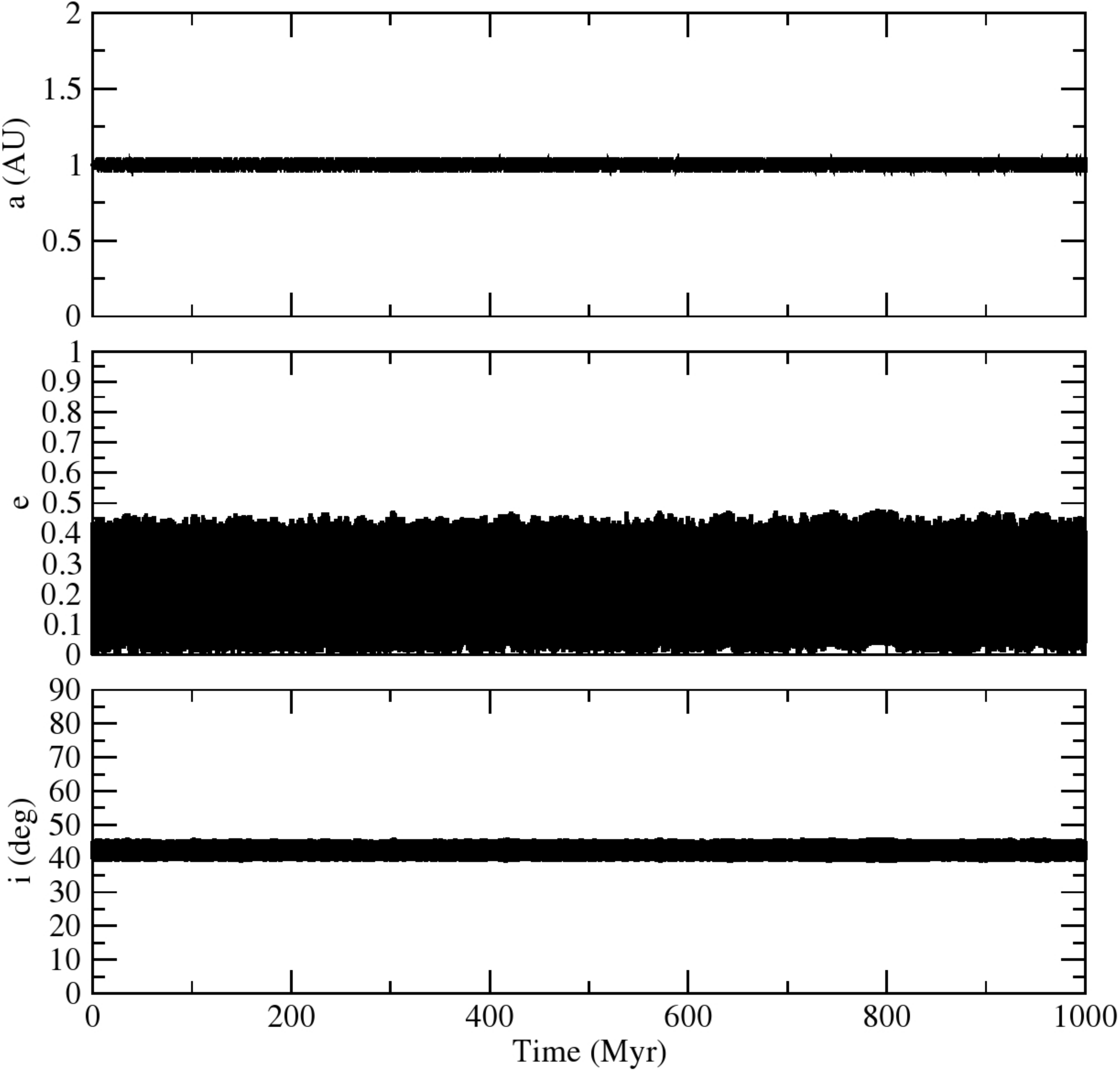}\\
\includegraphics[width=0.5\textwidth,clip=true,trim=0cm 0cm 0cm 0cm]{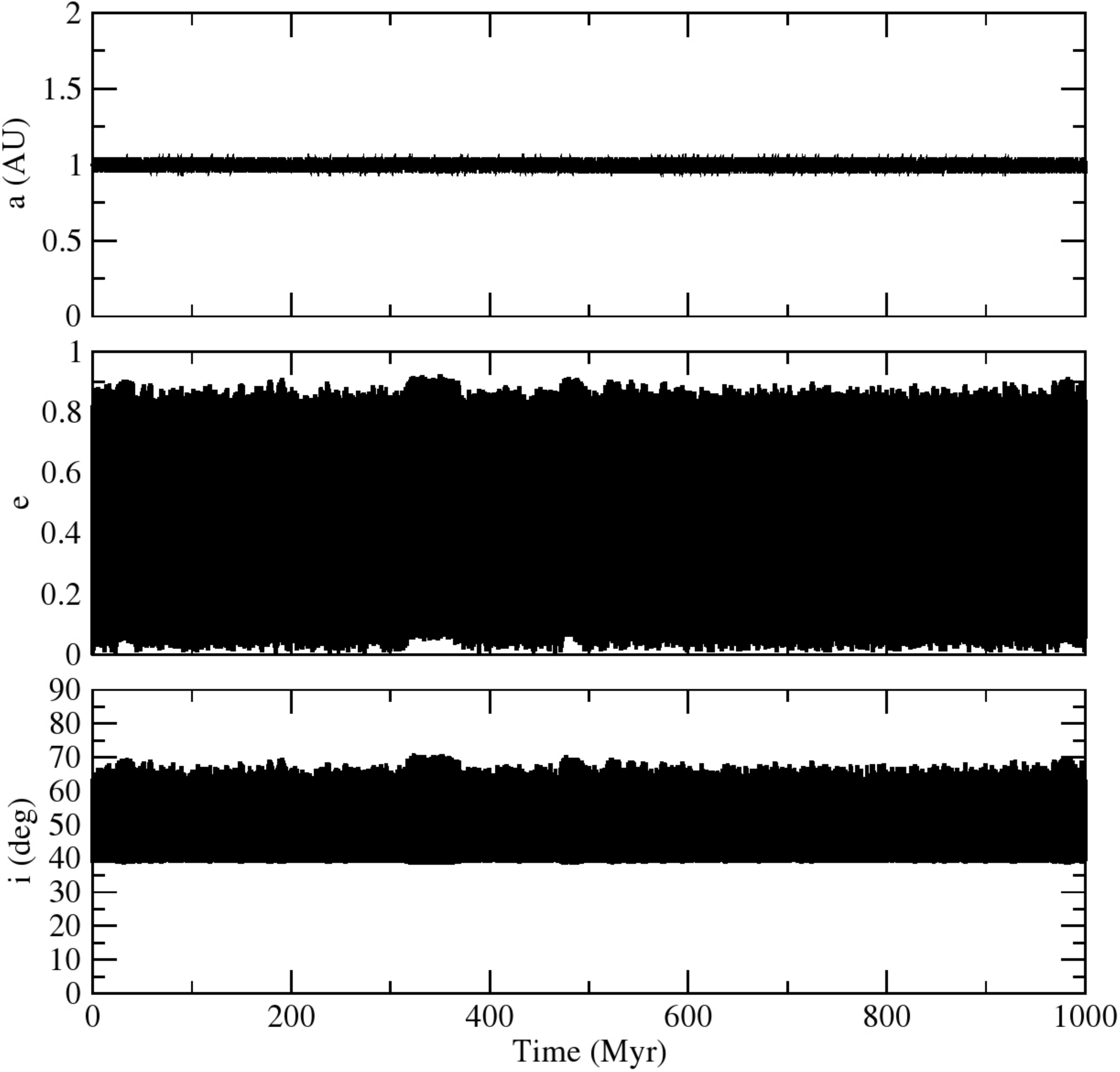}
\includegraphics[width=0.5\textwidth,clip=true,trim=0cm 0cm 0cm 0cm]{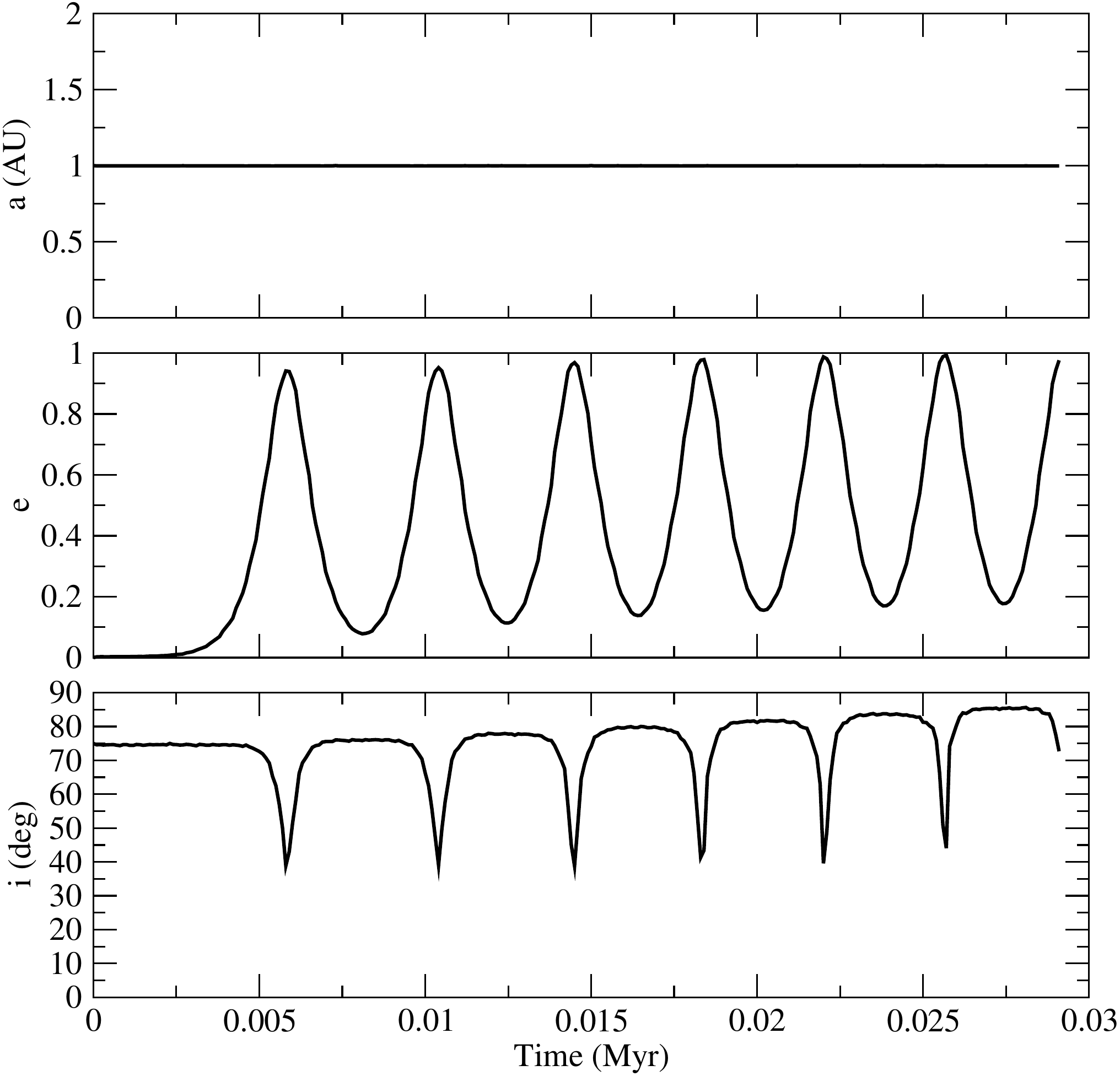}
\caption{The same as Figure 4, but with an initial semi-major axis of 1 AU.}
\end{figure*}

\begin{figure*}
\includegraphics[width=0.5\textwidth,clip=true,trim=0cm 0cm 0cm 0cm]{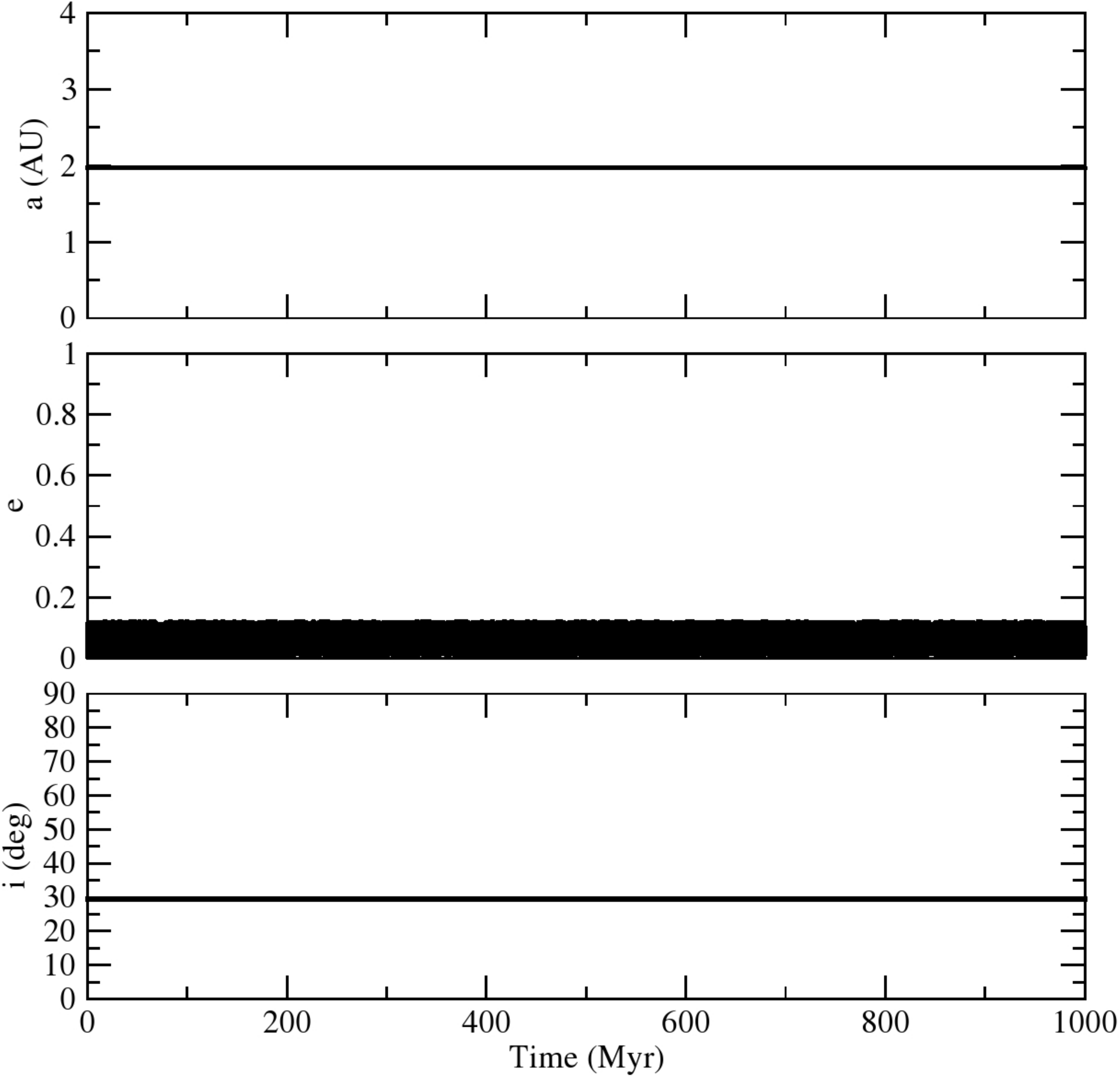}
\includegraphics[width=0.5\textwidth,clip=true,trim=0cm 0cm 0cm 0cm]{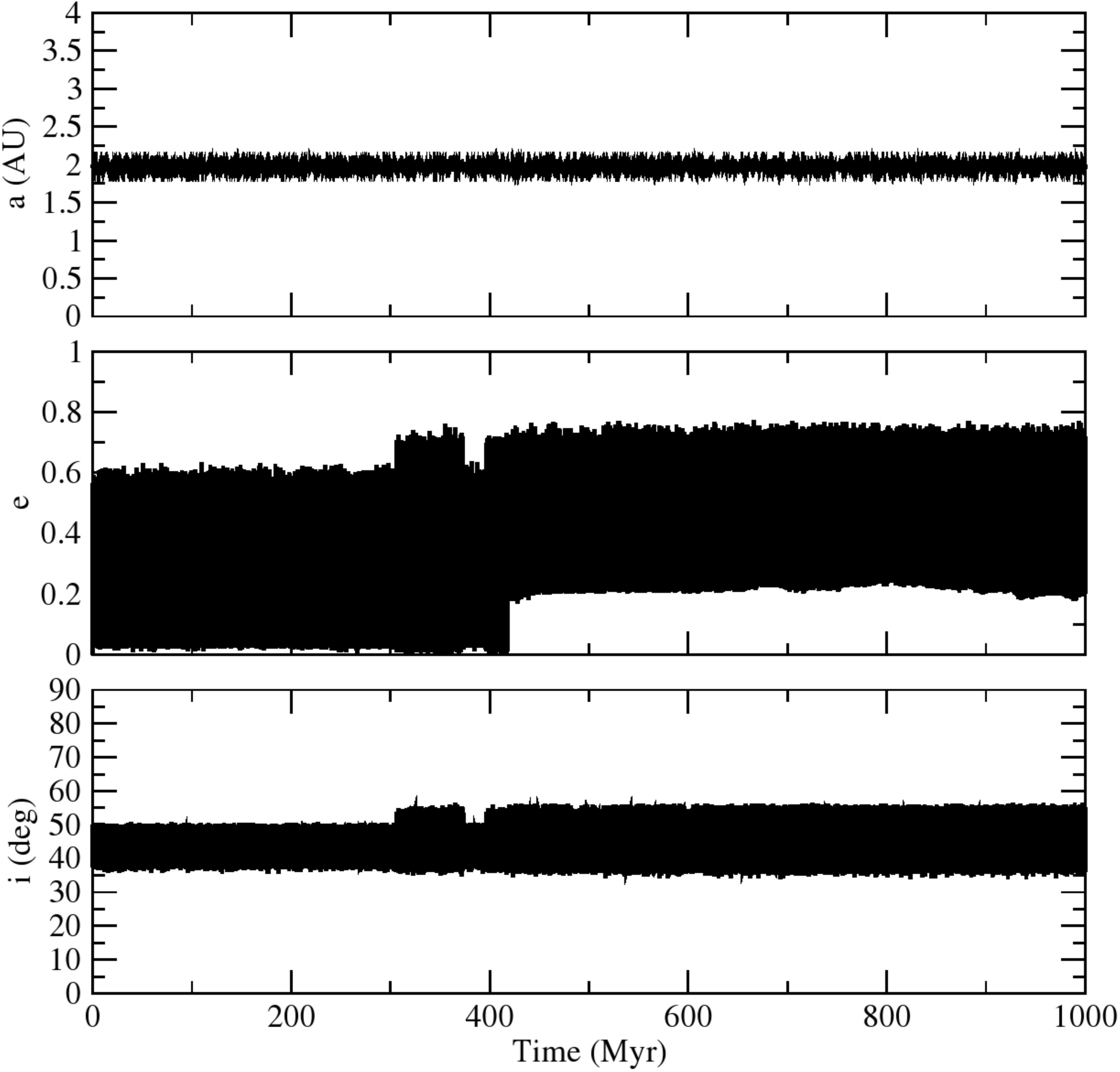}\\
\includegraphics[width=0.5\textwidth,clip=true,trim=0cm 0cm 0cm 0cm]{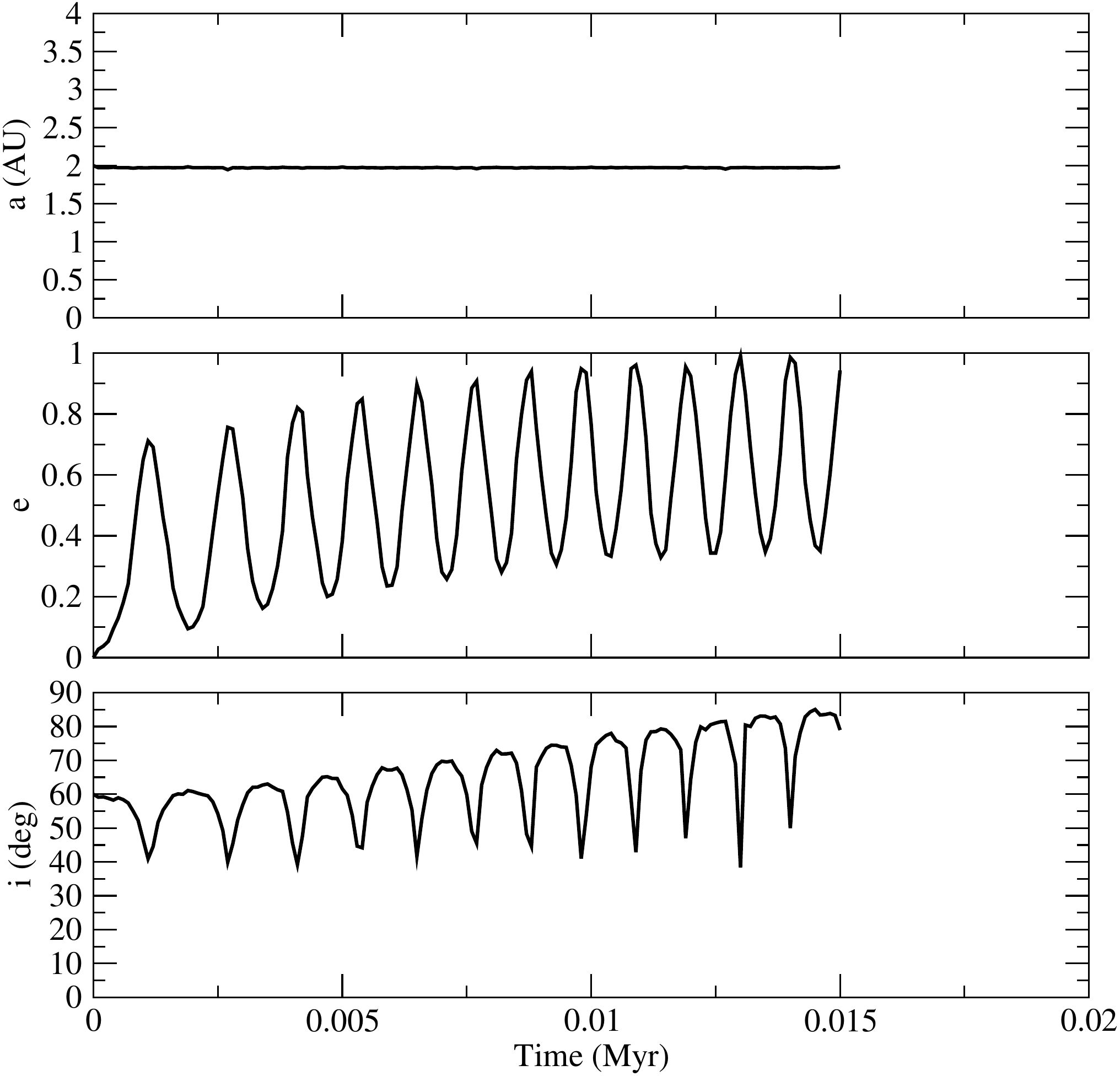}
\includegraphics[width=0.5\textwidth,clip=true,trim=0cm 0cm 0cm 0cm]{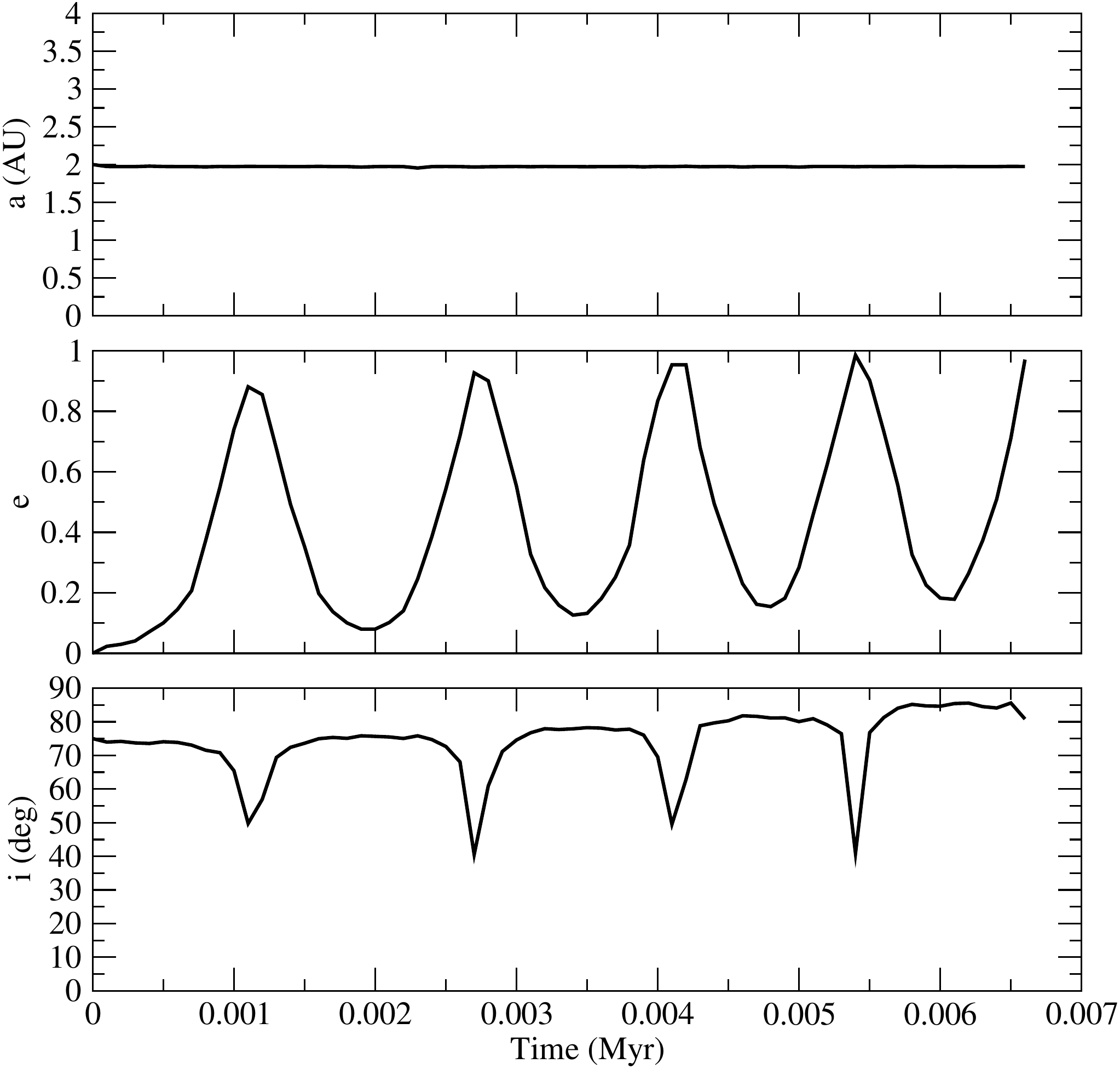}
\caption{The same as Figure 4, but with an initial semi-major axis of 2 AU.}
\end{figure*}

\clearpage

\begin{figure*}
\includegraphics[width=1.0\textwidth,clip=true,trim=0cm 0cm 0cm 0cm]{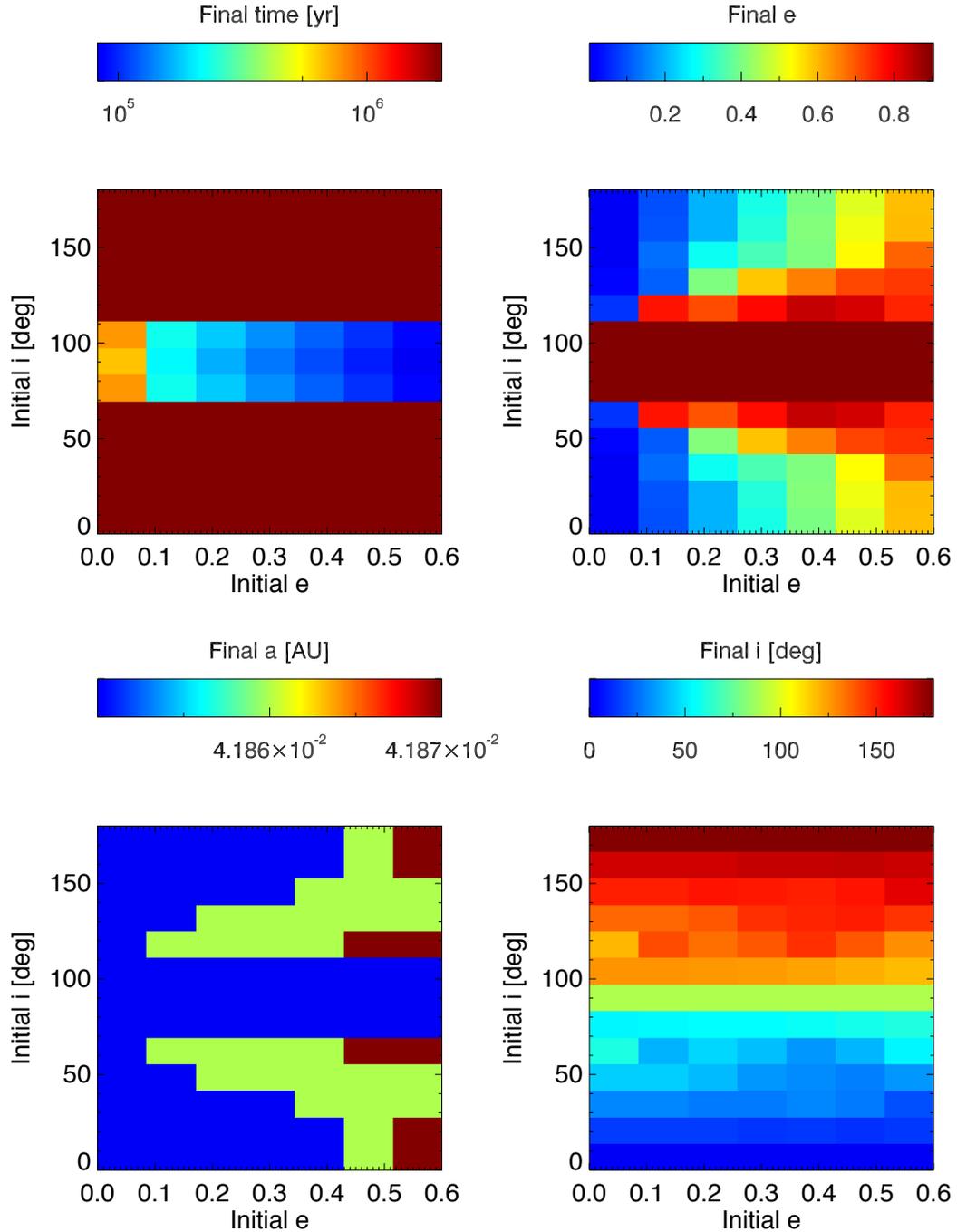}
\caption{Simulation results summary for a set of simulations ran without inclusion of tidal forces nor general relativistic precession (e.g. a basic N-body integration).  $\alpha$ Cen B b is initially placed at its current location of 0.042 AU from $\alpha$ Cen B, for a range of initial eccentricities and prograde and retrograde inclinations.  All simulations are carried out to 2 Myr, and a final time of $<$2 Myr indicates a planet ejection or collision with $\alpha$ Cen B.  Top left: the final integration time;  Top right: final eccentricity; Bottom left: final semi-major axis; Bottom right: final inclination.  This figure shows that without accounting for tidal forces and general relativistic precession, $\alpha$ Cen B b would be ejected by the Kozai mechanism for prograde inclinations of $>$60$^\circ$ within 2 Myr.}
\end{figure*}

\begin{figure*}
\includegraphics[width=1.0\textwidth,clip=true,trim=0cm 0cm 0cm 0cm]{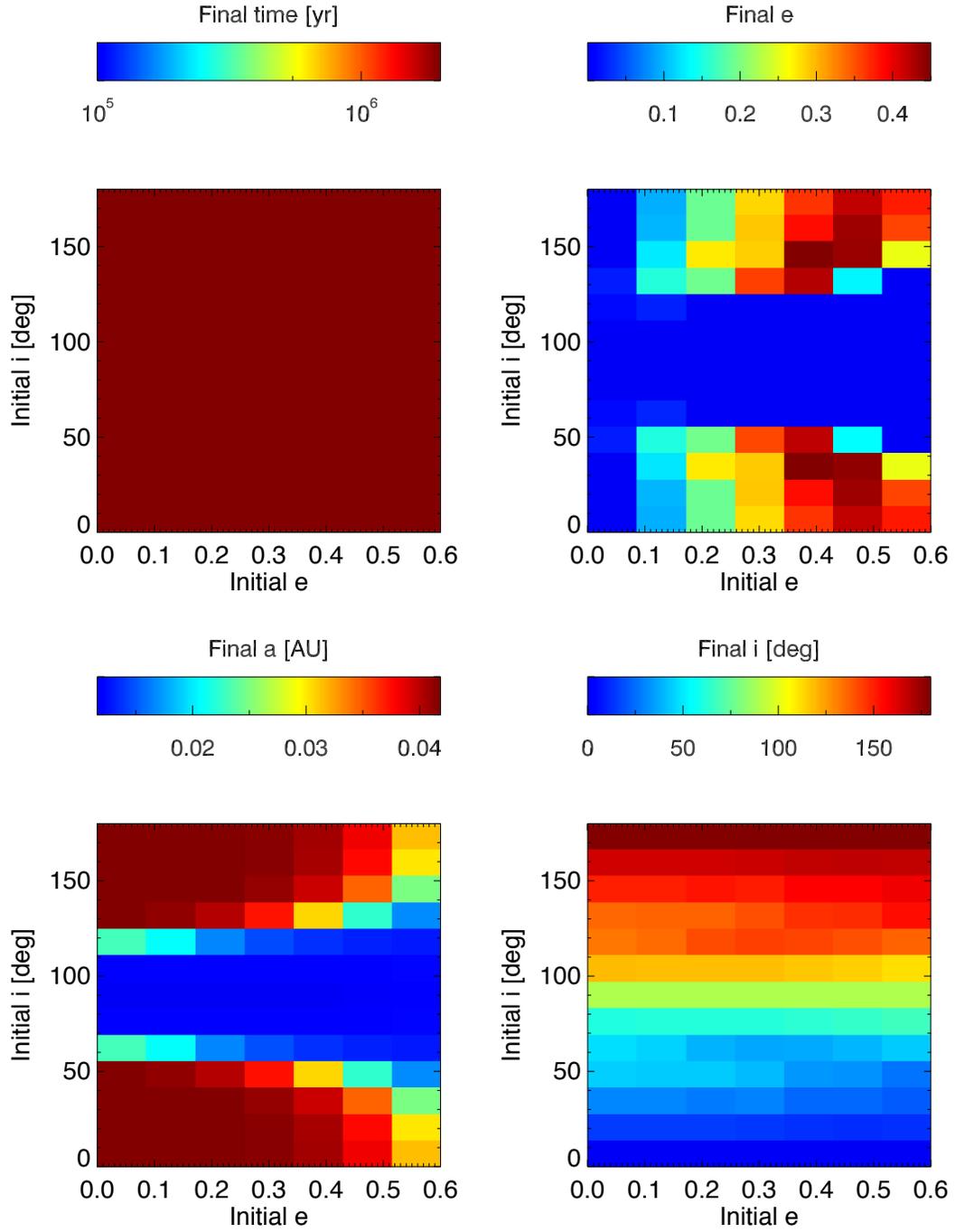}
\caption{The same as Figure 8, with tidal forces included in the simulations, but not general relativistic precession.  This figure shows that highly-inclined eccentric orbits are rapidly circularized within 2 Myr by tidal forces and the Kozai mechanism, but retain the initial inclinations.}
\end{figure*}

\begin{figure*}
\includegraphics[width=1.0\textwidth,clip=true,trim=0cm 0cm 0cm 0cm]{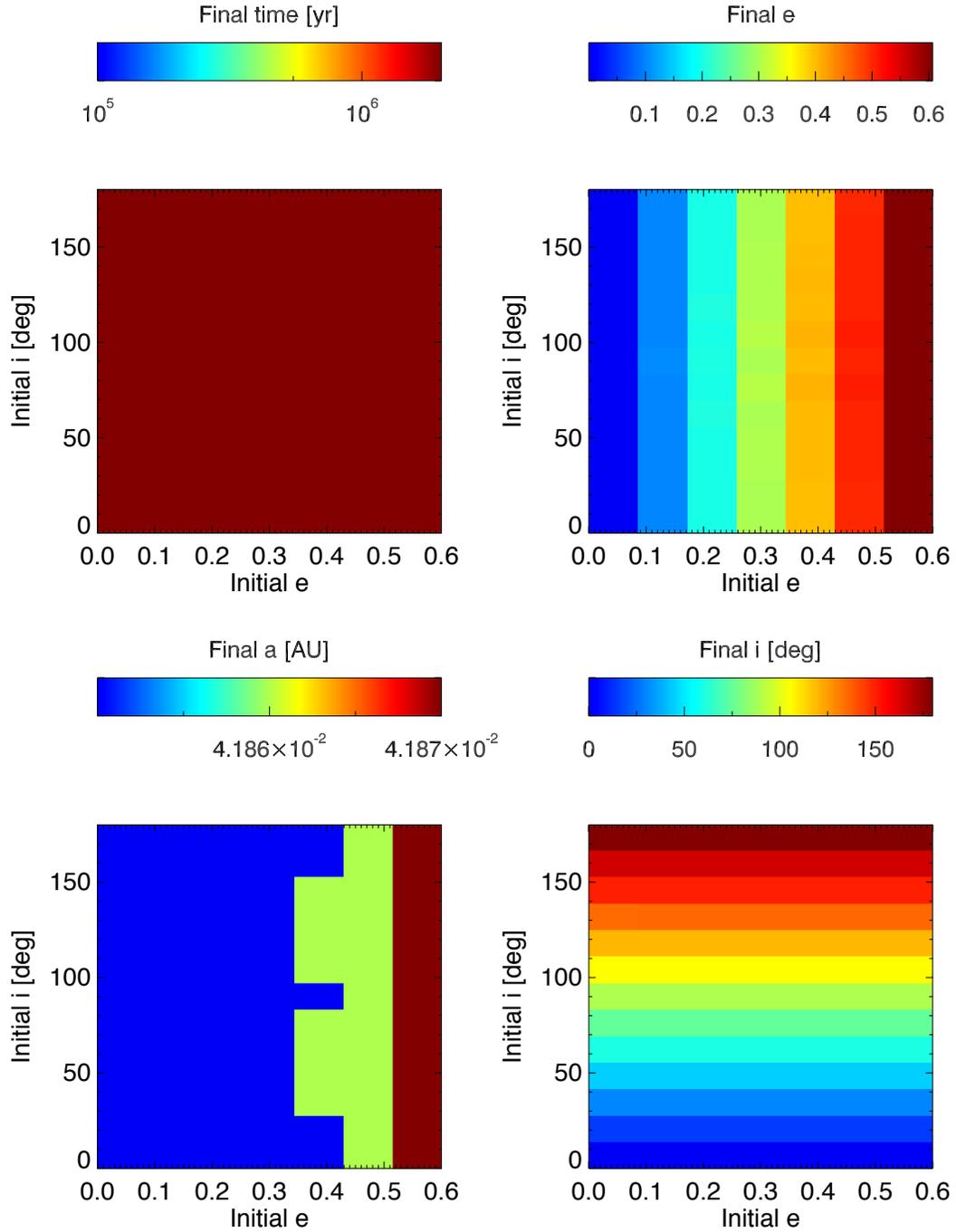}
\caption{The same as Figure 8, with general relativistic precession included in the simulations, but not tidal forces.  This figure shows that general relativistic precession dominates the orbital evolution of $\alpha$ Cen B b at its present location, mitigating the dynamical influence of the Kozai mechanism.}
\end{figure*}

\begin{figure*}
\includegraphics[width=1.0\textwidth,clip=true,trim=0cm 0cm 0cm 0cm]{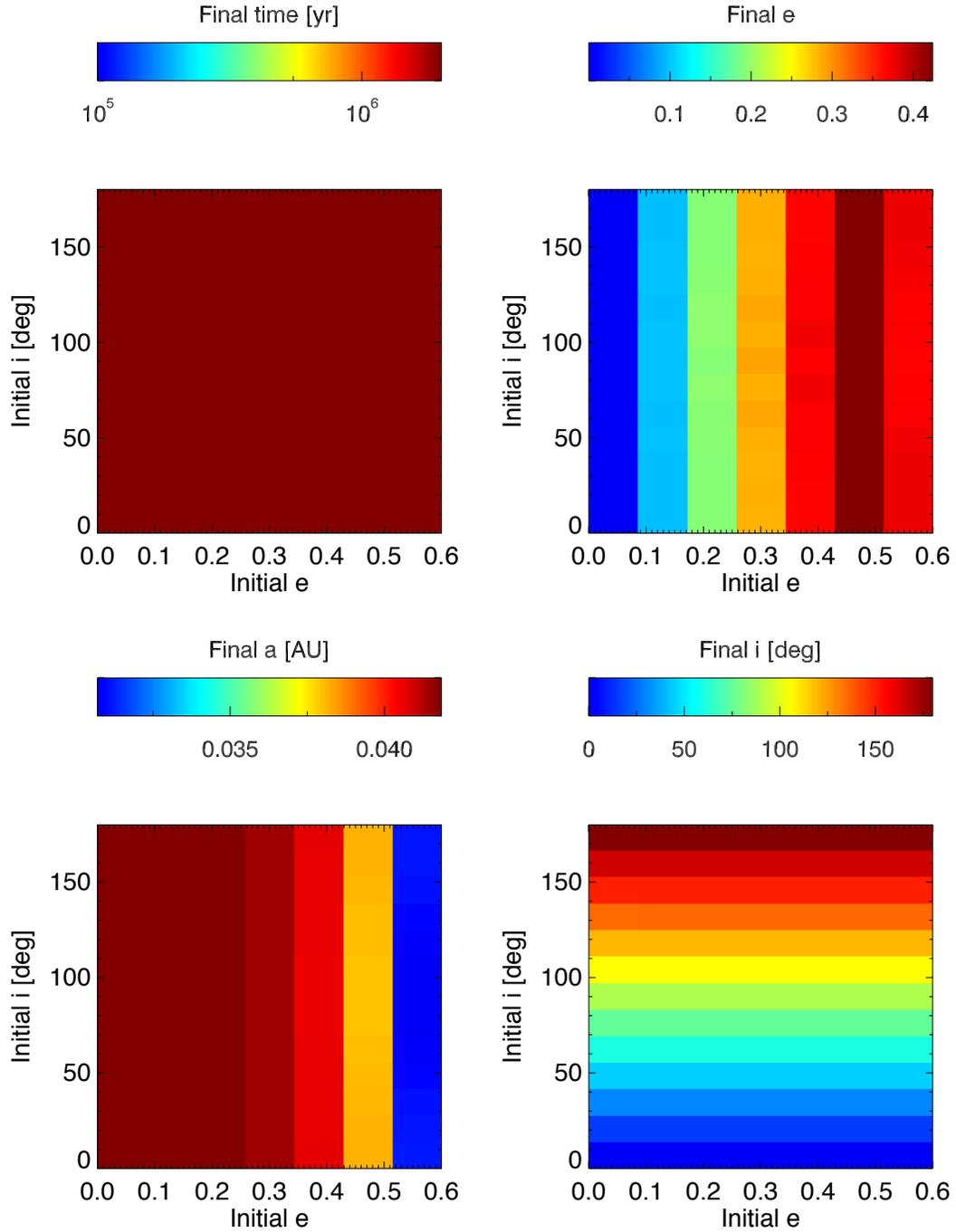}
\caption{The same as Figure 8, with general relativistic precession and tidal forces included in the simulations.  This figure shows that $\alpha$ Cen B b is dynamically stable at all possible prograde and retrograde inclinations at its present location for 2 Myr, and that the general relativistic precession has the effect of slowing down the tidal circularization of the orbit of $\alpha$ Cen B b at this semi-major axis.}
\end{figure*}

\begin{figure*}
\includegraphics[width=1.0\textwidth,clip=true,trim=0cm 0cm 0cm 0cm]{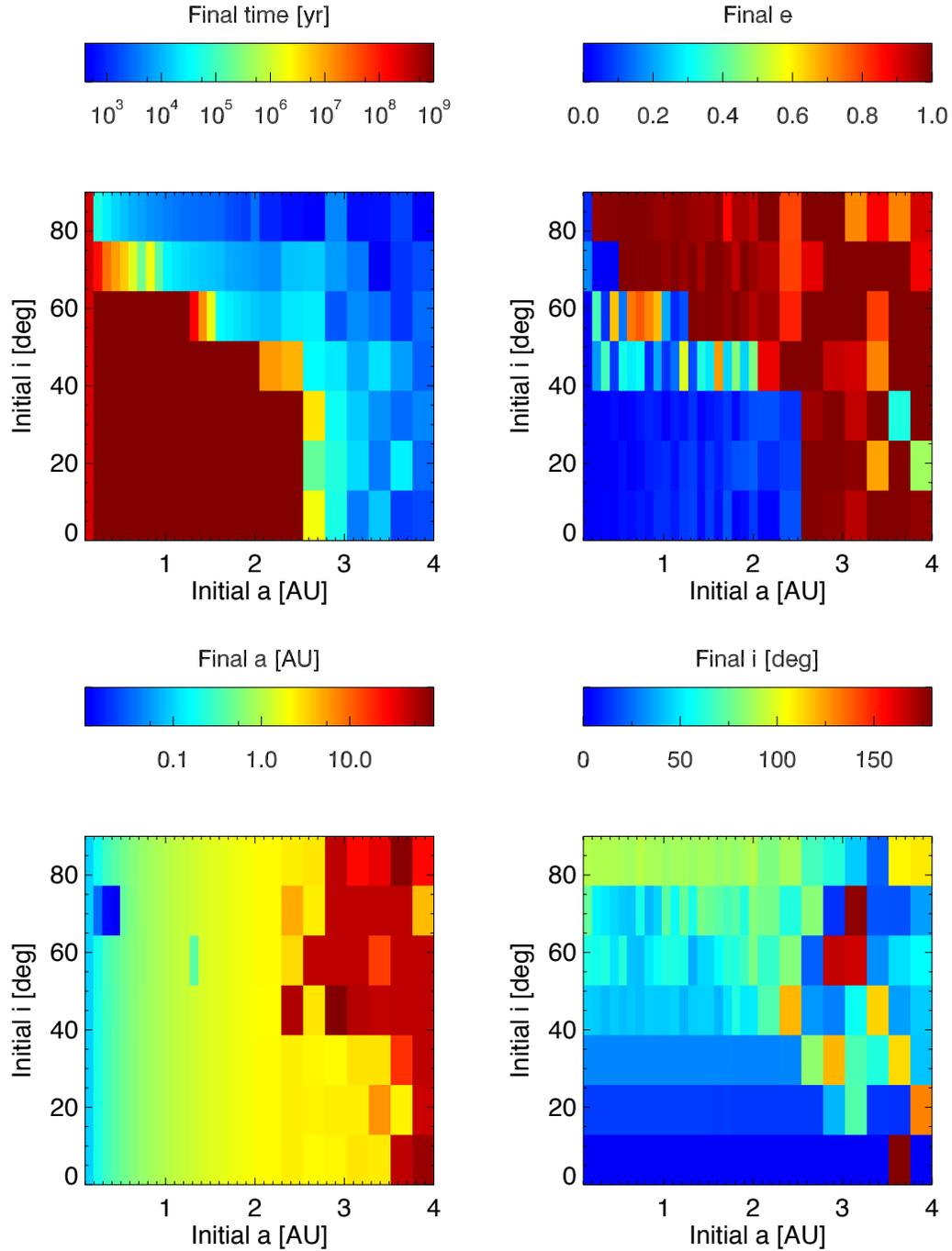}
\caption{The same as Figure 8, but now showing the simulations for a range of initial inclinations and semi-major axes, all with an initial eccentricity of zero, including both the tidal forces and general relativistic precession, and with simulations carried out for a duration of 1 Gyr (with the exception of simulations with a starting semi-major axis of 0.1 AU, which were halted after a duration of $\sim$250 Myr.  These plots show a clear stability and ejection regions for simulated planets as described in the text.}
\end{figure*}

\begin{figure*}
\includegraphics[width=1.0\textwidth,clip=true,trim=0cm 0cm 0cm 0cm]{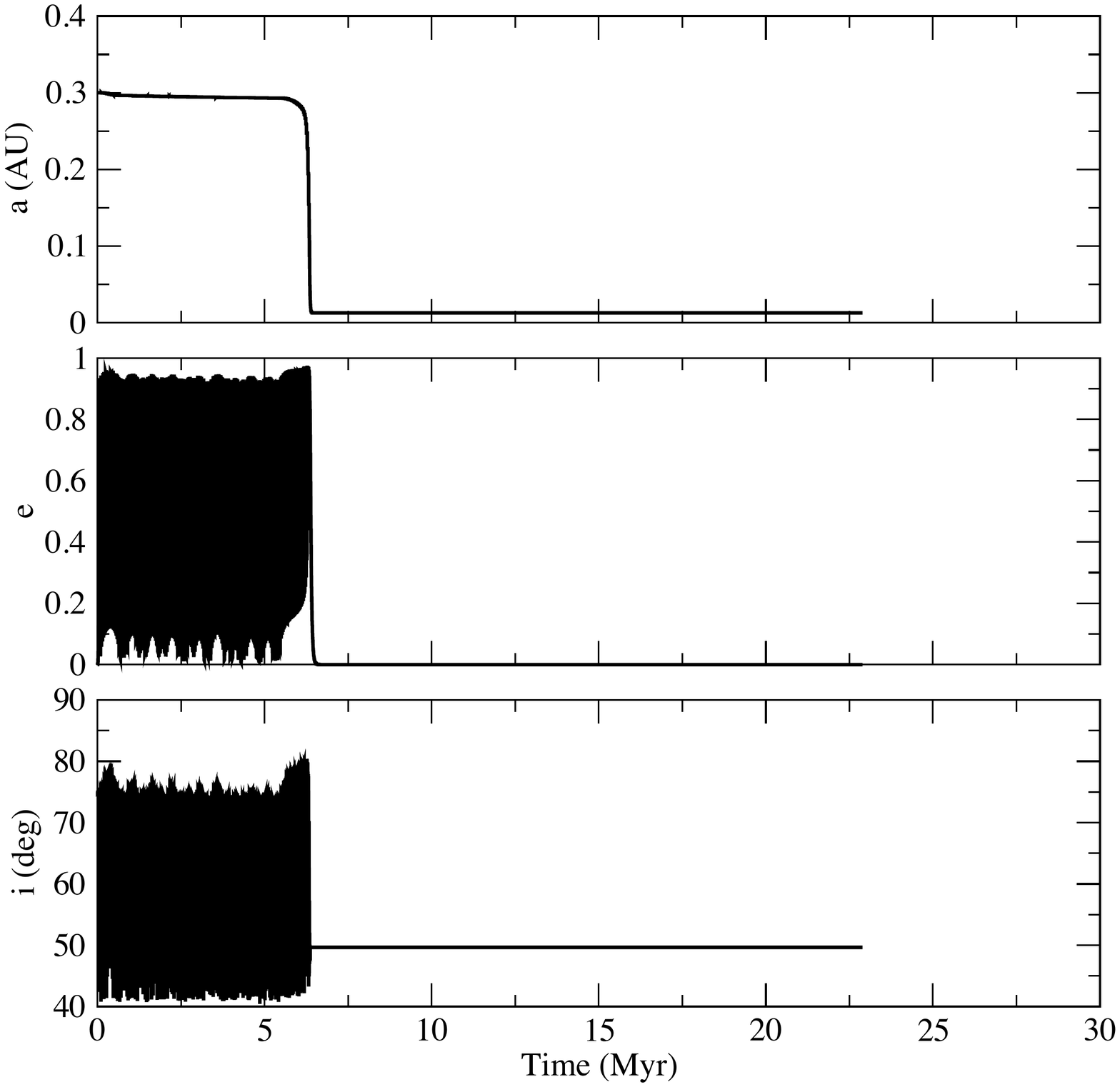}
\caption{An individual simulation of $\alpha$ Cen B b, with an initial semi-major axis of 0.3 AU and an initial inclination of 75$^\circ$ with respect to the AB orbital plane.  This simulation resulted in the rapid Kozai migration of the planet to $\sim$0.01, just exterior to the Roche radius of $\alpha$ Cen B within 100,000 yr. Data points are plotted every 100 years.}
\end{figure*}

\begin{figure*}
\includegraphics[width=1.0\textwidth,clip=true,trim=0cm 0cm 0cm 0cm]{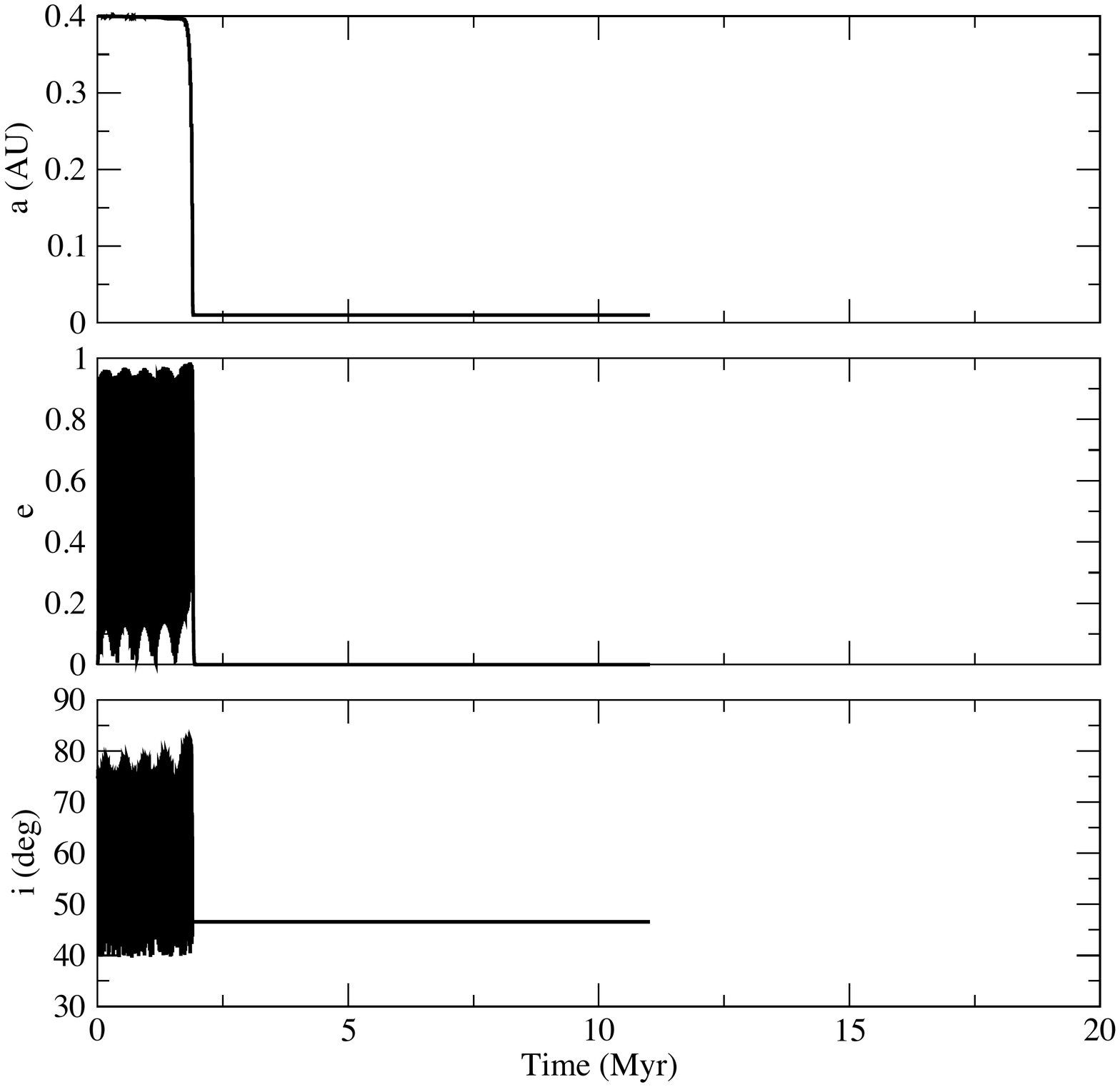}
\caption{An individual simulation of $\alpha$ Cen B b, with an initial semi-major axis of 0.4 AU and an initial inclination of 75$^\circ$ with respect to the AB orbital plane.  This simulation resulted in the rapid Kozai migration of the planet to $\sim$0.01, just exterior to the Roche radius of $\alpha$ Cen B within 100,000 yr. Data points are plotted every 100 years.}
\end{figure*}

\begin{figure*}
\includegraphics[width=1.0\textwidth,clip=true,trim=0cm 0cm 0cm 0cm]{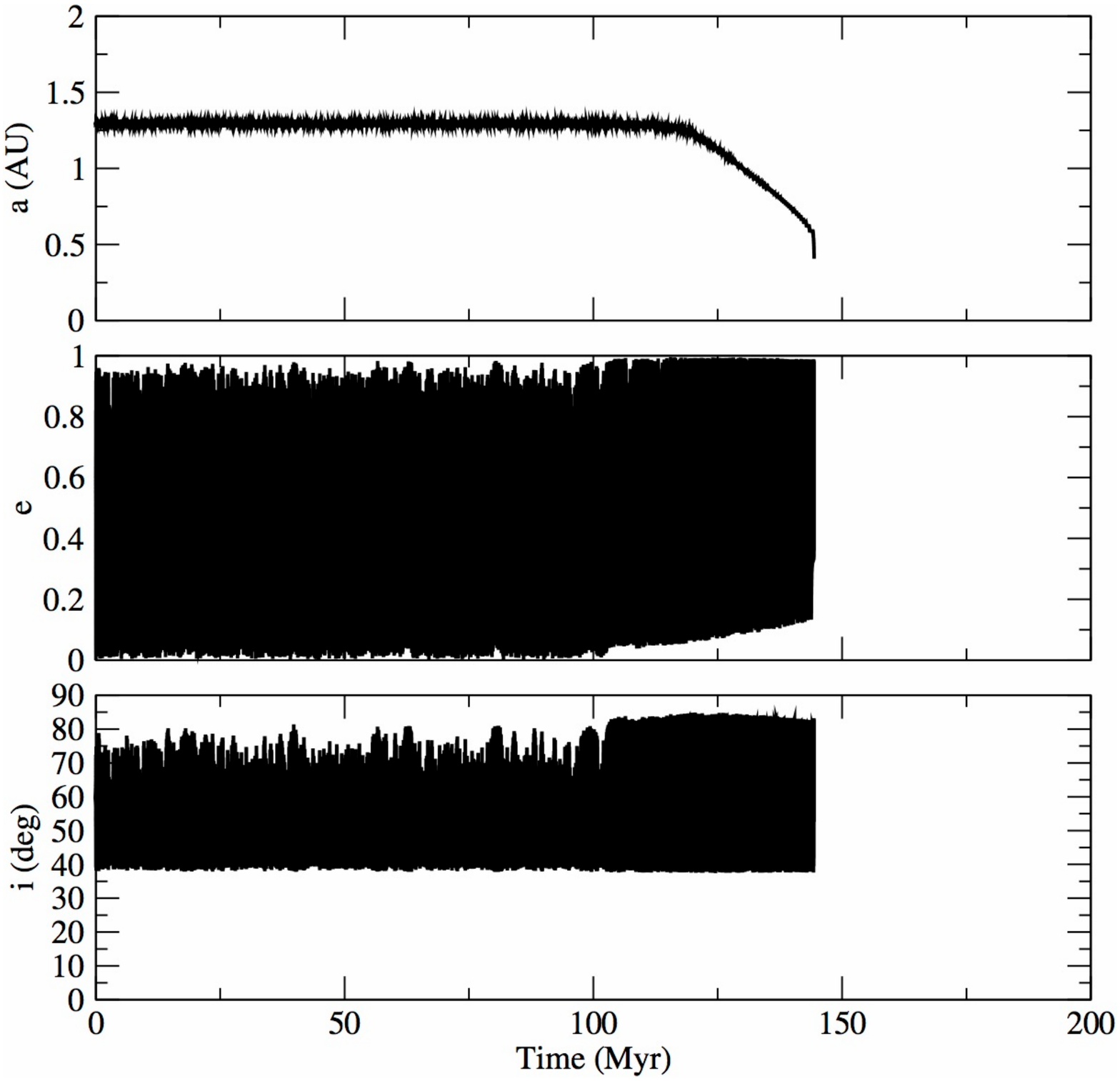}
\caption{An individual simulation of $\alpha$ Cen B b, with an initial semi-major axis of 1.3 AU and an initial inclination of 60$^\circ$ with respect to the AB orbital plane.  This simulation resulted in the initiation of a stable Kozai migration of the planet after $\sim$120 Myr to 0.6 AU after $\sim$145 Myr.    However, the planet then collided with $\alpha$ Cen B.  Data points are plotted every 100 yr.}
\end{figure*}

\section{Discussion and Constraints on the Orbital Inclination of $\alpha$ Cen B b from the Literature}

In this section, we discuss the results of our simulations, and combine them with existing literature to infer formation scenario dependent constraints on the present-day orbital inclination of $\alpha$ Cen B b.

\subsection{Simulation Implications}

The results of our simulation suggest that we cannot place any dynamical constraints on $\alpha$ Cen B b at its present location, with stable orbits found at all inclinations tested.  However, the current inclination of $\alpha$ Cen B b with respect to the AB binary orbital plane is also tied to the formation mechanism of the exoplanet.  $\alpha$ Cen B b could have formed in situ, migrated to its current location in a disk, migrated in resonance with another planet, or migrated via the Kozai mechanism or via planet-planet scattering among other possibilities.

We first consider whether it is feasible that  $\alpha$ Cen B b migrated via the Kozai mechanism to its present location.  It is clear from Figure 12 that a Kozai migration in this system requires a fine-tuned decreasing initial inclination as a function of increasing semi-major axis.  Higher initial inclinations are dynamically unstable, and lower inclinations are stable.   In our three simulations in which Kozai migration successfully completed, the final inclinations are 46$^\circ$--53$^\circ$ relative to the AB orbital plane and less than the initial inclinations.  Only one of our three migration simulations resulted in a circularized orbit exterior to the Roche radius of the star at a position comparable to the present semi-major axis of $\alpha$ Cen B b.  The migration in the other two simulations is so rapid that the planet possibly does not survive the tidal dissipation of orbital energy.   For the planet with an initial inclination of 60$^\circ$ and semi-major axis of 1.3 AU (Figure 15), a steady inward Kozai migration does start.  However, the process is clearly very fine-tuned to initial conditions, as the planet is destroyed halfway through the migration. Consequently, the Kozai migration is likely not a robust planet formation outcome for $\alpha$ Cen B b in this system from initial semi-major axes larger than $\sim$1 AU.

While the fragility of the Kozai migration mechanism may seen unexpected, \citet{wu2} does note that Kozai migration is not a common outcome of their simulations of HD 80606 b.  Further, the binary companion to HD 80606 in \citet{wu2} is located at 1100 AU, compared to $\sim$23 AU for $\alpha$ Cen A. The much shorter orbital time-scale for $\alpha$ Cen A, when compared to the tidal circularization and GR precession time-scales, may account for the fragility of the Kozai mechanism in our simulations.

Similarly, although not directly demonstrated in our simulations, planet-planet scattering events are unlikely to result in a significantly inclined orbit of $\alpha$ Cen B b with respect to the AB orbital plane at its present semi-major axis.  If $\alpha$ Cen B b was initially formed at a larger semi-major axis, and it received a significant orbital inclination boost from a planet scattering event, then the Kozai oscillation from $\alpha$ Cen A would dominate the dynamical evolution of $\alpha$ Cen B b thereafter.  Thus, $\alpha$ Cen B b would still require a fine-tuned inclination as a function of semi-major axis to avoid ejection or tidal disruption.  

We next turn to consider the formation of $alpha$ Cen B b in situ, or formation and then migration in a primordial disk with or without a hypothetical second planet in resonance.  

\subsection{Previous Planet Formation Simulations of the $\alpha$ Cen AB system}

Detailed numerical simulations of planet formation in binaries (Quintana et al. 2007, Fragner et al. 2011, Xie et al. 2011, Wu et al. 2007, Zhao et al. 2012), and in particular planet formation around $\alpha$ Cen A \& B (Rafikov \& Silsbee 2014, Andrade-Ines \& Michtchenko 2014, Guedes et al. 2008, Quintana et al. 2002, Lissauer et al. 2004, Quintana \& Lissauer 2006, Xie et al. 2010) give us the strongest constraints on the range of allowed inclinations for exoplanets in the $\alpha$ Cen system.  The salient points from this extensive list of references can be summarized as follows:

\begin{enumerate}
\item{Planet formation is more efficient around $\alpha$ Cen A compared to $\alpha$ Cen B.}

\item{Planet formation is less efficient for increasing misalignments, decreasing rapidly at inclinations of $\sim>25-45^{\circ}$ due to the Kozai mechanism.}

\item{Mutual orbital inclination decreases with decreasing orbital semi-major axis.}

\item{Planet formation of $\sim$2-5 exoplanets within 2 AU is feasible.}

\item{Short orbital period planets within $\sim$0.05 AU are often discarded due to the short dynamical time-scale w/r/t to numerical time-steps for computational efficiency, which is one factor that motivated our analysis.}

\item{Additional simulations of the $\alpha$ Cen system assume coplanarity in a 2D code and are thus not relevant to our discussion (Kley \& Nelson 2007,  Muller \& Kley 2012).}
\end{enumerate}

In particular, the results of oligarchic growth simulations carried to 200 Myr in \citet[][Table 1]{guedes} yield populations of planets with a standard deviation of inclinations of $\sim\pm8.6^{\circ}$, orbital semi-major axes of $\sim$0.2--1.8 AU, and eccentricities of $\sim$0.02-0.35 from an initial set of co-aligned planetesimals.  Additionally, Quintana et al. 2002 specifically investigated oligarchic growth simulations carried to 400 Myr for a planetesimal disk initially inclined w/r/t to the AB orbital plane.  The outcome of the simulations for planet formation around $\alpha$ Cen A (Figures 4, 8, 9, Quintana et al. 2002) and $\alpha$ Cen B (Figure 10) generally produce planets inside of $\sim$0.5 AU with inclinations of $<20^{\circ}$, regardless of the initial planetesimal disk inclination w/r/t to the AB binary orbital plane.  For planets exterior to $\sim$1 AU, Quintana et al. 2002 find that the proto-planet can retain their initial mis-alignment.  Finally, Zhao et al. 2012 presented an analysis of the inclination evolution of an exoplanet in the presence of a primordial gas disk in a binary system and concluded that the inclination will remain small as well.  

From these literature simulations, we conclude that $\alpha$ Cen B b could have likely formed with an initial inclination $\pm20^{\circ}$ w/r/t to the AB binary orbital plane, which is less than the critical Kozai angle of 39.2$^\circ$.  Whether  $\alpha$ Cen B b migrated in a disk with or without a resonance planetary companion, or formed in situ at its present location, these literature simulations strongly suggest that $\alpha$ Cen B b is not misaligned with the AB orbital plane by more than 20$^\circ$.  However, \citet[]{quintana2} start their simulation with planetesimals exterior to 0.36 AU.  Thus, for semi-major axes inside of this value, we are extrapolating our conclusion about the evolution of proto-planets inclinations. Future detailed studies of planetesimal evolution interior to 0.36 AU are warranted.  

$\alpha$ Cen C is currently far enough away ($\sim$15,000 AU) from the AB binary to be ignored as having any current dynamical influence on $\alpha$ Cen B b.  However, at some point during the early formation of the $\alpha$ Cen system, assuming the C component is bound, C may have had a closer approach to the AB system in a fashion sufficient to warp or disturb the circumsecondary disk or young protoplanets around the B component.  This could have resulted in a misalignment, disruption, or migration of the B circumsecondary disk / protoplanets, but such a misalignment would not have been likely to survive the dynamical influence of $\alpha$ Cen A in our simulations and literature simulations.

\subsection{$\alpha$ Cen B and its Stellar Spin Axis Alignment}

We next consider if we can infer any constraints on the orbital inclination of $\alpha$ Cen B b from the stellar-spin axis alignment of $\alpha$ Cen B. The spin of $\alpha$ Cen A is observed to be aligned with the AB orbital plane, as inferred from projected rotational velocity combined with the observed rotation period and radius measurements obtained from asteroseismology and interferometry (Bazot et al. 2007, Kervella 2003).  However, the slower rotation period of $\alpha$ Cen B ($\sim$36-42 days; Dewarf et al. 2010, Jay et al. 1997, Buccino \& Mauas 2008), combined with the low vsini of 1.1$\pm$0.8 km/s (Saar \& Osten 1997), precludes a useful constraint on the inclination of the stellar spin axis from R$\sim$10$^6$ spectroscopy (Frutiger et al. 2005).  Given the observed rotation period and stellar radius, the expected rotational velocity is $\sim$1 km/s for $\alpha$ Cen B, which is consistent with the observed v sin i.  

The best observational constraint on the inclination of the spin axis of $\alpha$ Cen B comes from Dumusque (2014).   Through the modeling of simultaneous radial velocity and photometric observations of $\alpha$ Cen B, Dumusque (2014) derives an inclination on the sky of the stellar spin axis of $\alpha$ Cen B of 45$^{+9}_{-19}$ degrees.  With an unknown orientation on the sky, this corresponds to a minimum misalignment of $>$20$^\circ$ at 2-$\sigma$ with the orbital plane of the AB binary.

We can make a related analysis by comparing the stellar jitter activity level of $\alpha$ Cen B to the Sun.  D12 reported the observation of differential rotation for $\alpha$ Cen B from the radial velocity jitter induced by the rotational modulation of starspots.  The 2008-2011 radial velocity observations in D12 yield periods of 39.76, 37.80 and 36.71 days in 2009,2010, and 2011 respectively.  These observations span the minimum to maximum in the $\sim$8.8 yr stellar activity cycle reported for $\alpha$ Cen B (DeWarf et al. 2010, Ayres 2009).  The rotation period evolution is consistent with $\alpha$ Cen B exhibiting a Sun-like ``butterfly diagram'' evolution of starspots from latitudes of $\sim\pm$30$^{\circ}$ to the stellar equator from epochs of minimum to maximum activity.  Thus, we can constrain the spin axis of $\alpha$ Cen B to be $\sim>30^{\circ}$ deviant from an axis normal to the sky.  Otherwise, the spots would be visible at all rotational phases in 2009, and the projected radial velocity rotational modulation would likely not be observed at the detected amplitude.  This is consistent with the measurement in Dumusque (2014), but as noted in Dumusque (2014) the rotation periods reported are susceptible to error because of the harmonic fitting to the radial velocity time-series.

We can take this line of inquiry one step further.  D12 derives radial velocity r.m.s. in the 2008-2011 seasons of 1.18, 1.50, 2.19 and 2.15 m/s respectively, corresponding to an increase in quadrature of $\sim$1.8 m/s in \textit{projected} radial velocity jitter from the increased stellar activity from 2008 to 2011.  The SunÕs expected radial velocity jitter due to rotational modulation is $\sim$0.4 m/s (Makarov et al. 2009). This value is a factor of $\sim$4 below the observed jitter for $\alpha$ Cen B despite the slower rotational velocity of $\alpha$ Cen B compared to the Sun -- $\sim$1.1 km/s for $\alpha$ Cen B vs. $\sim$1.6 km/s for the Sun (Pavlenko et al. 2012).    Further, D12 measures activity levels of $Log R^\prime_{HK}$ = -4.99, -4.94, -4.89 and -4.90 in 2008-2011 respectively, compared to the solar mean activity level of -4.90 (Mamajek et al. 2008).  In other words, the activity level of $\alpha$ Cen B is comparable to that of the Sun.  If we assume $\alpha$ Cen B to have a spot frequency, size, and temperature contrast similar to that of the Sun, the projected radial velocity jitter of $\alpha$ Cen B is difficult to reconcile with the estimated jitter of the Sun without the spin axis of $\alpha$ Cen B being nearly perpendicular to the normal to the plane of the sky (e.g. $\sim>60^{\circ}$).   Instead, $\alpha$ Cen B must have larger spots or larger spot temperature contrast to account for the factor of $\sim$3 needed to reconcile the observed RV jitter and rotation period relative to the Sun, which likely can be attributed to the differences in spectral type.  

Finally, if the axis of stellar spin was perpendicular to the normal to the sky, and aligned with the direction of orbital motion rather than the orbital plane, this would be consistent with the arguments presented thus far.  However, a noticeable cycle in the activity of $\alpha$ Cen B could be apparent over the course of the 80 year orbital period of the AB binary, and this is not observed with detailed studies of activity dating back a few decades (Flannery \& Ayres 1978, Dewarf et al. 2010).  While this scenario is not expressly discounted, we conclude from the above arguments that the stellar spin axis of $\alpha$ Cen B is likely $\sim>30^{\circ}$ deviant from the normal to the plane of the sky, with the most likely value of 45$^\circ$ coming from Dumusque (2014) implying a misalignment of $>$20$^\circ$ with the AB orbital plane.  However, the misalignment is likely not significantly larger -- e.g. $\sim<$45$^\circ$ misalignment -- lest we run into difficulty to accounting for the observed differential rotation and jitter amplitude, and lack of activity modulation over the $\sim$80 year binary orbit. Thus, unlike $\alpha$ Cen A, $\alpha$ Cen B is likely $\sim$20--45$^\circ$ misaligned with the AB orbital plane.

\subsection{Binary Stellar Spin Alignment}

Barring direct observational constraints of the stellar spin alignment of $\alpha$ Cen B, we can invoke spin-orbit alignment measurements of young binaries of comparable separations to $\alpha$ Cen AB in Hale (1994), Howe \& Clarke (2009), and at larger separations in Jensen et al. (2004) and Monin et al. (2006), to estimate the likely spin-orbit alignment of $\alpha$ Cen B.  These studies identify that at separations of $\sim<$100 AU, the spins of stars in a young binary are typically aligned to within $\sim10-30^{\circ}$ of the orbital plane.  Additionally, observations of young low mass stellar binaries indicate that circum-primary and circm-secondary disks are usually aligned with the orbit of the binary as well (Prato et al. 2007, Monin et al. 2006, Watson et al. 2011, Wheelwright et al. 2011).  These observations are supported by binary disk modeling of systems including eccentric systems like $\alpha$ Cen (Pichardo et al. 2005).   Thus, this suggests on an ensemble basis, although it is not conclusively demonstrated through direct observation, that the spin of $\alpha$ Cen B is aligned with the AB orbital plane to within $\sim30^{\circ}$.  Such an alignment is consistent with the observed rotational modulation and radial velocity jitter amplitude of $\alpha$ Cen B presented in D12 and marginally consistent with Dumusque (2014), given the arguments presented herein thus far.

On the other hand, however, Skemer et al. (2008), Jensen \& Akeson (2014) and Roccatagliata et al. (2011) present evidence for mis-aligned disks in the triple system T Tauri, the binary system HK Tauri, and the binary Haro 6-10 respectively.  Jensen et al. (2004) notes that compact triples can increase the odds of spin-orbit misalignment.  Thus, if Proxima Centauri was closer to $\alpha$ Cen AB earlier in the evolution of this triple system, it could have warped the circum-secondary disk around $\alpha$ Cen B into misalignment. 

\subsection{Spin-Orbit Alignment of $\alpha$ Cen B b}

In ${\S}$4.3 and 4.4, we have made the argument that the spin of $\alpha$ Cen B is aligned with the orbital plane of the AB binary with an angle of $\sim$20-45$^{\circ}$ from direct observational constraints, and from less conclusive comparative studies of binary stars with and without circumstellar disks.   We next consider if the orbital plane of the exoplanet $\alpha$ Cen B b is aligned with the stellar spin of $\alpha$ Cen B.  This might be expected for an exoplanet forming in situ or migrating in an aligned proto-planetary disk. Observations of solitary transiting exoplanets in short orbital periods around single stars indeed demonstrate that the exoplanet orbit and stellar spin are well-aligned for older systems with effective temperatures $<$6000 K and exoplanet masses $>$0.2 M$_J$ (Albrecht et al. 2012, Winn et al 2010).  Additionally, for multiple, compact terrestrial exoplanet systems, co-alignment within a few degrees is the norm (Fang et al. 2012).  However, there are a number of short-period systems that are mis-aligned, and this is thought to be due to migration of the exoplanets via planet--planet scattering or the Kozai mechanism.  Further, a number of studies have predicted that exoplanets in binary systems are more likely to be mis-aligned (Parker \& Goodwin 2009, Wu et al 2007, Xie et al. 2011).

Tidal interactions of a close-in exoplanet, even if initially misaligned, can be re-aligned with the stellar spin-axis within the tidal circularization time-scale (Winn et al. 2010).   Given the age of $\alpha$ Cen of $\sim$5 Gyr, one could assume that the system has had adequate time to re-align $\alpha$ Cen B b to the stellar spin axis of $\alpha$ Ceb B via the action of stellar tides, regardless of any initial misalignment.  However, the mass of $\alpha$ Cen B b is orders of magnitude smaller than 0.2 M$_J$, and thus the re-alignment time-scale is orders of magnitude longer (Albrecht et al. 2012).  It is likely that $\alpha$ Cen B b is not massive enough to excite stellar tides to induce the re-alignment on a time-scale that is shorter than the age of the $\alpha$ Cen system.  Indeed, HAT-P-11 b is one of the lowest mass exoplanets with a measured Rossiter-McLaughlin effect at $\sim$26 M$_\oplus$, and it is an outlier that is misaligned with the stellar spin axis of its host star (Winn et al. 2010).  More observations of the Rossiter-McLaughlin effect for single, short-period terrestrial mass planets are needed to determine if orbital re-alignment is likely or rare.   Thus, we cannot conclusively argue that an initially mis-aligned $\alpha$ Cen B b was re-aligned with its host stellar spin axis.  However, we can assert that if the initial planet-forming disk around $\alpha$ Cen B was aligned with the spin axis of $\alpha$ Cen B and not the AB orbital plane, and if $\alpha$ Cen B b did not undergo Kozai migration, then $\alpha$ Cen B b will have likely retained that primordial spin-orbit alignment (Fang et al. 2012).

\section{Conclusions}

The constraints that we can place on the inclination of $\alpha$ Cen B b and consequently its true mass are reliant on considerations of the formation mechanism of the planet. Without any consideration of the planet formation mechanism, our dynamical simulations show that the inclination of $\alpha$ Cen B b in its present location and consequently its true mass are unconstrained.  However, by considering possible formation scenarios, we can place a useful constraint on the inclination of the planet:

\begin{enumerate}

\item{If the planet migrated via the Kozai mechanism or planet-planet scattering to its present location, even if this is an unlikely outcome from our simulations, then the current inclination of $\alpha$ Cen B b is likely $<$55$^\circ$ mis-aligned with the AB orbital plane, corresponding to an angle of $<$65$^\circ$ with respect to our line of sight.  While this may seem like a very weak constraint, this corresponds to an upper mass limit of $<$2.7 M$_\oplus$ ($<$3.3 M$_\oplus$ taking the 3-$\sigma$ upper limit to the m$sin$i from D12), with a transit probability of 15\%.}

\item{If instead $\alpha$ Cen B b formed in situ, or migrated in a proto-planetary disk with or without an additional planet in orbital resonance, then previous literature dynamical simulations suggest that $\alpha$ Cen B b is $<$20$^\circ$ mis-aligned with the AB orbital plane, corresponding to an angle of $<$30$^\circ$ with respect to our line of site, and a mass of $<$1.3 M$_\oplus$ ($<$1.6 M$_\oplus$ taking the 3-$\sigma$ upper limit to the m$sin$i) and a transit probability of 30\%.}

\item{Finally, if we consider the recent result in Dumusque et al. (2014) and our interpretation that $\alpha$ Cen B is misaligned with the AB orbital plane by $\sim$20--45$^\circ$, and if we instead assume that the orbit of $\alpha$ Cen B b is aligned with the spin axis of $\alpha$ Cen B from an initial primordial disk alignment with the spin of $\alpha$ Cen B, then we can estimate the inclination of $\alpha$ Cen B b to also be $\sim$10--55$^\circ$ inclined with respect to our line of site, with a negligible transit probability and a mass of 1.14--2.7 M$_\oplus$.}

\end{enumerate}

Thus, independent of the formation scenario, we can conclude that $\alpha$ Cen B b likely possesses a mass $<$2.7 M$_\oplus$ orbiting a star in the nearest stellar system to the Sun ($<$3.3 M$_\oplus$ taking the 3-$\sigma$ upper limit to the m$sin$i from D12).  This puts the planet in the range of masses expected for terrestrial planets (Marcy et al. 2014).  

In addition to confirming the terrestrial nature of $\alpha$ Cen B b, future determination of the composition and bulk density of $\alpha$ Cen B b will help discern amongst the formation mechanisms presented in this paper.  Unfortunately, given the recently reported stellar spin-orbit misalignment of the $\alpha$ Cen B with the AB orbital plane, the transit probability is possibly low and requires a space-based photometer.  Regardless, transits are still worth excluding.  Future direct imaging surveys will be challenged by the maximum projected separation of $\sim$33 milli-arcseconds of $\alpha$ Cen B b and $<$1 milli-arcseconds astrometric stellar reflex motion of $\alpha$ Cen B, and the light from $\alpha$ Cen A.  Time-resolved interferometric measurements of the spot rotation of $\alpha$ Cen B b could help confirm the reported stellar spin angle of $\alpha$ Cen B.   

Our conclusion relies on the assumption that the radial velocity detection of $\alpha$ Cen B b is secure.  Future observational radial velocity constraints on the $m$sin$i$ are warranted to confirm the planet detection.  Finally, our analysis may be applicable to the planets found in the binary systems such as HD 41004, $\gamma$ Ceph, and Gliese 86 (Muller \& Kley 2012, Akeson et al. 2013). 

We thank the anonymous referee for a thoughtful review that improved the clarity and presentation of this manuscript.  The authors would like to acknowledge Thayne Currie, Dave Latham and David Ciardi for their encouragement in writing this paper.  This research has made use of the NASA Exoplanet Archive, which is operated by the California Institute of Technology, under contract with the National Aeronautics and Space Administration under the Exoplanet Exploration Program.  P. Plavchan would like to acknowledge support from a NASA JPL Research and Technology Development program, and the Missouri SpaceGrant Consortium.

\end{document}